\documentclass[review]{elsarticle}

\usepackage{epsfig}
\usepackage[font=small,labelfont=bf]{caption}
\usepackage{amssymb,amsmath,graphicx,latexsym}

\usepackage{graphicx}
\usepackage{caption}
\usepackage{subcaption}

\usepackage{bm}
\usepackage[table]{xcolor}
 \usepackage{url}

\usepackage{booktabs}
\usepackage[utf8]{inputenc}
\usepackage{lmodern}

\usepackage[font=small,skip=0pt]{caption}
\usepackage[T1]{fontenc}
\usepackage{rotating}
\usepackage{pdflscape}
\usepackage{tabulary}
\usepackage{array}
\usepackage{booktabs}
\usepackage{appendix}
\usepackage{color, colortbl} 
\usepackage[table]{xcolor}
\usepackage[first=0,last=9]{lcg}

\usepackage{graphicx}
\usepackage{caption}
\usepackage{subcaption}

\usepackage[section]{placeins}
\usepackage{diagbox}
\usepackage{graphics}

\newtheorem{theorem}{Theorem}[section]

\newtheorem{definition}{Definition}[section]

\usepackage[
top    = 3.50cm,
bottom = 3.50cm,
left   = 3.50cm,
right  = 4.00cm
]{geometry}

\usepackage{lipsum}
\makeatletter
\def\ps@pprintTitle{%
 \let\@oddhead\@empty
 \let\@evenhead\@empty
 \def\@oddfoot{}%
 \let\@evenfoot\@oddfoot}
\makeatother

\begin{document}


\title{\bf Modeling Count Data via Copulas 
}

  \cortext[cor]{Corresponding author}

 \author{Hadi Safari-Katesari}
 \author{S. Yaser Samadi \corref{cor}}
 \ead{ysamadi@siu.edu }
 
 \author{Samira Zaroudi \corref{co}}

  \address{Department of Mathematics, Southern Illinois University,  Carbondale IL 62901, USA }

\begin{abstract}
Copula models have been widely used to model the dependence between continuous random variables, but modeling count data via copulas has recently become popular in the statistics literature. Spearman's rho is an appropriate and effective tool to measure the degree of dependence between two random variables.  In this paper, we derived the population version of Spearman's rho correlation via copulas when both random variables are discrete. The closed-form expressions of the Spearman correlation are obtained for some copulas of simple structure such as Archimedean copulas with different marginal distributions. We derive the upper bound and the lower bound of the Spearman's rho for Bernoulli random variables.   Then, the proposed  Spearman's rho correlations are compared with their corresponding Kendall's tau values. We characterize the functional relationship between these two measures of dependence  in some special cases. An extensive simulation study is conducted to demonstrate the validity of our theoretical results.  Finally, we propose a bivariate copula regression model to analyze the count data of a  \emph{cervical cancer} dataset.
\end{abstract}
 \begin{keyword}
Spearman's rho, Copula, Bivariate measure of association, Concordance Discordance Dependence.

\end{keyword}

 \maketitle
\date{ }

\section{Introduction}
 Measuring  association and dependence between random variables has always been a main concern of statisticians.  In dependency theory, correlation is defined as a measure of dependence or statistical relationship between two random variables.  The correlation and association between random variables can be captured using different measures. Many of these measures are based on the concept of concordance and discordance probabilities when discrete random variables are involved.
     We say two random variables are concordant if large values of one variable tend to be correlated with large values of the other and small values of one with small values of the other (see Nelsen, 2006). On the other hand,  two random variables are discordant if large values of one variable tend to be associated  with small values of the other and vice versa.    A variety of concordance-discordance based measures have been proposed in the literature, for instance  Kendall's $\tau$ proposed by Kendall (1945), Spearman's rho proposed by Spearman (1904), Blomqvist's $\beta$ proposed by Blomqvist (1950), Goodman's $\gamma$ proposed by Goodman and Kruskal (1954), Kendall's $\tau_ b$ proposed by Agresti (1996), Stuart's $\tau_ c$ proposed by Stuart (1953), and the Somers' $\Delta$ proposed by Somers (1962). In this paper, we focus on the two most important and commonly used concordance-based dependence measure of associations, i.e.,  Spearman's  rho and  Kendall's  tau for discrete random variables.

   It is well known that the dependence measures derived through copulas are more informative than the classical measures.
 Copula models have been extensively used to measure the dependence between continuous random variables, e.g., Nelsen (2006) has studied a wide range of important copula-based dependence measures, particularly  Spearman's rho when the marginal distributions are continuous. Due to the positive probability of ties  in discontinuous cases, the copula-based dependence measures constructed for continuous random variables cannot be used for discrete cases.
   Several authors such as Tchen (1980), and Scarsini (1984) have tried to formulate and measure the dependency  between discrete  random variables in the class of concordance measures. 
    Moreover, Sklar (1959) has shown that a multivariate copula with discrete marginal distributions does not have a unique copula representation.
  Also, Genest and Ne\'{s}lehov\'{a} (2007) demonstrated that the copula for count data with discrete marginal distributions  is not identifiable, and this problem occurs when
    one of the marginal distributions is discontinuous. More details of the identifiability issue of the copula can be found in Genest and Ne\'{s}lehov\'{a} (2007) and Trivedi and Zimmer (2017).  In the discrete context, one of the biggest barriers is the non-uniqueness of the associated copulas. Different authors (e.g.,  Mesfioui and Tajar, 2005; Denuit and Lambert, 2005; and Ne\'{s}lehov\'{a},  2007)  have addressed this problem by  proposing different transformations to derive a continuous extension of discrete random variables.

     Mesfioui and Tajar (2005),  Denuit and Lambert (2005), Nikoloulopoulos (2007), among others,  proposed the population version of  Kendall's tau, and derived it by using copula function when the marginal distributions are discrete.
     Quessy (2009) considered multivariate generalization of Kendall's tau and Spearman's rho for multivariate ordinal data, and proposed several test statistics for testing independence of ordinal random variables.   Mesfioui and Quessy (2010) introduced multivariate extensions of Kendall's tau, Spearman's rho, and Spearman's footrule for discontinuous random variables.  Genest et al. (2013) obtained asymptotic variance of Spearman’s rho for multivariate count data. Genest et al. (2014) considered the empirical multilinear copula process for multivariate  count data, and  established the asymptotic distribution of the empirical process. Liu et al. (2018) defined a partial and conditional  Spearman's rho based on concordance and discordance probabilities.  Moreover,  Genest et al. (2019) proposed   consistent and distribution-free tests for testing the mutual independence of arbitrary random variables.
     Loaiza-Maya and Smith (2019) proposed the Spearman's rho  for stationary ordinal-valued time series data.

 In this paper, we focus on a discrete setting and use a similar  procedure as that presented in Mesfioui and Tajar (2005), Denuit and Lambert (2005), and Nikoloupolous and Karlis (2009) to obtain the population version of Spearman's rho when the margins are discrete random variables  based on concordance and discordance probabilities.   Particularly, we focus  on deriving the Spearman's rho for the discrete margins  by taking into account the principle of continuity proposed by Schriever (1986) and Denuit and Lambert (2005). For brevity and simplicity of notation, we use the letters  ``$C"$, $``D"$, and $``T"$  to denote  ``concordance'', ``discordance'', and ``tie'', respectively.
  The main property of the concordance family in discrete cases is that the probability of tie plays an important role such that $P(C)+P(D)+P(T)=1$. Notice that,  in continuous cases, the probability of tie is zero.  As a byproduct of these results, we compare Spearman's rho and Kendall's tau by plotting them over different values of the corresponding parameter and compare their behaviors with different types of copulas with the same  margins. In particular,  the functional relationship between  these two dependence measures  are characterized by numerical features    when the margins are Binomial, Negative Binomial, and Poisson.

The rest of the paper is organized as follows.
In Section 2,  the classical notations and fundamental definitions  used in the sequel are introduced.
The population version of Spearman's rho via copulas when both random variables are discrete is proposed in Section 3. In particular, the upper
and lower bounds of Spearman’s rho with Bernoulli margins are derived.
In Section 4, numerical analyses are conducted to compare the behaviors of  Spearman's rho and Kendall's tau obtained  by some well-known Archimedean family of copulas, such as the Frank, Gumbel and Clayton copulas.  Poisson and Bernoulli variables are used as marginal distributions. Their lower and upper bounds are tested  numerically to validate our theoretical results. Moreover, an extensive simulation study is performed to demonstrate the validity of our theoretical results. In Section 5, we analyze a real data on \emph{Cervical Cancer}, modeled as a negative binomial for  both margins.   All of the proofs are presented in  the Appendix. 

  \section{Spearman's rho for Count Data}\label{sec2}

The main purpose  of this paper is to find the population version of  Spearman's rho for discrete random variables by using copula functions and  based on concordance and discordance measures. Therefore, it is appropriate to review these terms which will be used to obtain the population version of  Spearman's rho for count data. Moreover, the continuation principle and the procedure  of the continuous extension of discrete margins is used that preserves the concordance order, and as a result it preserves  Spearman’s rho.

\subsection{Concordance and Discordance}
 Similar to Kendall's tau, Spearman's rho dependence measure is built on concordance and discordance probabilities. Two random variables are concordant if large values of one variable are associated  with large values of the other variable, and vice versa (Nelsen, 2006). Similarly, two random variables are disconcordant if large values of one variable tended to have small values of the other variable.  The probability of these two concepts and the probability of tie are defined in Definition \ref{def.1}  below.
 \begin{definition}\label{def.1}
Let $(X_{1},Y_{1})$ and $(X_{2},Y_{2})$  be  two independent realizations  from the joint distribution of  $(X,Y)$. Then, the probability of  ``concordance'',  ``discordance'', and ``tie''  are, respectively, defined as follows
\begin{align}\label{eq1}
  P(C)&=P\left[(X_{1}-X_{2})(Y_{1}-Y_{2})>0\right],  \\\label{eq2}
P(D)&=P[(X_{1}-X_{2})(Y_{1}-Y_{2})<0],  \\\label{eq3}
  P(T)&=P[X_{1}=X_{2}~~ or ~~Y_{1}=Y_{2}]  .
\end{align}
\end{definition}

Notice that, when marginal distributions are continuous, the probability of tie, $P(T)$,  is zero.    However,  this is not the case when the margins are discrete and therefore the probability of tie should be taken into account. 

\subsection{Copulas with Discrete Margins}
Copulas have become one of the most important tools to model and measure nonlinear dependence structure between random variables.
Unlike the continuous case, copulas with discrete margins are not unique (Sklar, 1959).  
\begin{definition}(Nelsen, 2006))
A two-dimensional copula function $\mathcal{C}(u,v)$  is a function  defined from the entire unit square to the unit interval  with the following properties:
\begin{enumerate}
\item
$\mathcal{C}(u,0)=\mathcal{C}(0,v)=0$        for all, $ u,v \in[0,1]$,
\item
$\mathcal{C}(u,1)=u,~~~\mathcal{C}(1,v)=v$        for all, $ u,v \in[0,1]$,
\item
$\mathcal{C}(u_1, v_1)-\mathcal{C}(u_2, v_1)-\mathcal{C}(u_1, v_2)+\mathcal{C}(u_2, v_2)\geq0$ ~for all,~ $ u_1,u_2,v_1,v_2 \in[0,1]$,~ if~ $ u_2\geq u_1$,$v_2\geq v_1$

\end{enumerate}
\end{definition}
Sklar (1959) showed  that any bivariate cumulative distribution function (CDF), e.g., $F_{X,Y}$ can be represented  as a function of its marginal CDFs, $F_X$ and $F_Y$, by using a two-dimensional copula function $\mathcal{C}(.,.)$, that is
\begin{equation}\label{eq4}
F_{X,Y} (x,y)=P(X\leq x,Y\leq y)=\mathcal{C}(F_X(x),F_Y(y)).
\end{equation}
Notice that, the copula function $\mathcal{C}(\cdot, \cdot)$ in Eq (\ref{eq4}) is unique if $F_X$ and $F_Y$  are continuous, however, when the marginal distributions are discrete, then the copula function $\mathcal{C}(\cdot, \cdot)$   is not unique.

 There are a few  drawbacks when marginal distributions are discontinuous. For instance, based on Sklar's theorem,  the copula function is not unique (identifiable) in the discrete case except on the range of the marginal distributions.  Moreover, it can be shown  that the range of Spearman's rho for  discrete random variables is narrower than  $[-1,1]$.  Nevertheless, the dependency parameter of the copula function can still demonstrate the dependency between the marginal variables.  For more details see Genest and Ne\'{s}lehov\'{a} (2007).
\subsection{Spearman's rho}
 Similar to Kendall's tau, Spearman's rho is  one of the  fundamental concepts of dependency and  mathematically is defined as follows.
Let $(X_{1},Y_{1})$, $(X_{2},Y_{2})$, and $(X_{3},Y_{3})$ be three   independent realizations  from the joint distribution of  $(X,Y)$; then, Spearman's rho is defined as (see Nelsen, 2006)
\begin{align}\label{eq5}
\begin{split}
  \rho^S(X,Y)&=3\left(P(C)-P(D)\right)\\
             &= 3\big(P((X_1-X_2)(Y_1-Y_3)>0)-P((X_1-X_2)(Y_1-Y_3)<0)\big).
\end{split}
\end{align}
 If $X$ and $Y$ are continuous random variables, then it can be shown that
\begin{align}\label{eq6}
\rho^S(X,Y)=12\int_{0}^{1}\int_{0}^{1}\mathcal{C}(u,v)dudv-3,
\end{align}
where $\mathcal{C}(\cdot, \cdot)$ is a copula function. However, when $X$ and $Y$ are discrete random variables, then the probability of tie is positive and we have   $P(C)+P(D)+P(T)=1$ . Therefore, the definition of Spearman's rho can be rewritten as follows
\begin{align}\label{eq7}
\begin{split}
\rho^S (X,Y)=&3\left(P(C)-P(D)\right) \\
        =&3\left( 2P(C)-1+P(T)\right)  \\
        =&6\bigg[ P\big((X_{1} -X_{2} )(Y_{1} -Y_{3} )>0\big)\bigg] -3+3P(X_{1} =X_{2} ~ or ~ Y_{1} =Y_{3}).
\end{split}
\end{align}
Note that, $X_{2}$ and $Y_{3}$ are independent in Eq (\ref{eq7}). In  Section \ref{sec3}, we will show that when the marginal distributions are discontinuous,   Spearman's rho has a narrower range than $[-1,1]$. This is because, in discontinuous cases, the probability of tie is positive. 
More details of the  drawbacks and limitations of Spearman's rho for dependent count data can be found in 
Park and Shin (1998), Mari and Kortz (2001), 
 and Madsen and Birkes  (2013).
\subsection{Continuation Principle for Discrete Variables }

Due to non-uniqueness of the copula for discontinuous random variables, it is very difficult to work with the original discontinuous margins. However, the continuous extension of discrete margins can be used if the desired properties persist under continuous extension. That is, we make discontinuous margins continuous by adding a perturbation taking values between zero to one.

Assume $X$ is a discrete random variable with probability mass function (pmf) $p_i=P(X=i),  i\in Z$. Notice that, since strictly increasing transformations of marginal distributions do not change the Spearman's rho (see  Mesfioui and Tajar, 2005), therefore, without loss of generality, we assume that  $X$ takes its values in  $Z$. 
Mesfioui and Tajar (2005) introduced  the following transformation in order to transform a  discrete random variable $X$  into  a continuous random variable $X^{*}$
\begin{equation}\label{eq8}
X^{*}=X+U,
\end{equation}
where $U$ is a continuous   random variable on $[0,1]$, which is  independent of $X$. Then, we say $X$ is continued by $U$.  
 Some mathematical properties of the discrete concordance measures have been investigated by Mesfioui and Tajar (2005). Similar to Denuit and Lambert (2005) that showed continuous extension  preserves  Kendall's tau, we  prove  that  the continuous extension also preserves  Spearman's rho. To this end,  assume $(X_1,Y_1) $,  $ (X_2,Y_2) $ and $(X_3,Y_3)$ are  three  independent copies of $(X,Y)$. Moreover, assume
\begin{enumerate}
\item[(i)] for $i=1,2,3$,  $X_{i}$ and $Y_{i}$ are continued by $U_{i}$ and $V_{i}$, respectively;
\item[(ii)]
$U_{1}, U_{2}, U_{3}, V_{1}, V_{2}, V_{3}$ are independent and continuous random variables on $[0,1]$;
\item[(iii)]
$U_{1}$, $U_{2}$, and $U_{3}$ ($V_{1}$,  $V_{2}$, and $V_{3}$) have the same distribution.
\end{enumerate}
Then, we have
\begin{align*}
P^{*}(C)=&P[(X_{1}^{*}-X_{2}^{*})(Y_{1}^{*}-Y_{3}^{*})>0]\\
=&P[(X_{1}+U_{1}-X_{2}-U_{2})(Y_{1}+V_{1}-Y_{3}-V_{3})>0]\\
=&P[X_{1}=X_{2},Y_{1}=Y_{3}]P[(U_{1}-U_{2})(V_{1}-V_{3})>0]\\
&+P[X_{1}=X_{2},Y_{1}>Y_{3}]P[U_{1}-U_{2}>0]+P[X_{1}=X_{2},Y_{1}<Y_{3}]P[U_{1}-U_{2}<0]\\
&+P[X_{1}>X_{2},Y_{1}=Y_{3}]P[V_{1}-V_{3}>0]+P[X_{1}<X_{2},Y_{1}=Y_{3}]P[V_{1}-V_{3}<0]\\
&+P[(X_{1}-X_{2})(Y_{1}-Y_{3})>0].
\end{align*}
Since  $U_{1}-U_{2}$ and $V_{1}-V_{3}$ are continuous random variables with symmetric density functions  around zero, we have
\begin{equation}\label{ }
   P[U_{1}-U_{2}>0]=P[V_{1}-V_{3}>0]=P[U_{1}-U_{2}<0]=P[V_{1}-V_{3}<0]=\frac{1}{2}.
\end{equation}
Note that, in the special case when $U_{i}$ and $V_{i}$  are uniformly distributed on $(0,1)$, then $U_{1}-U_{2}$  and $V_{1}-V_{3}$  have the  Triangle distribution$[-1,1,0]$,  which is a symmetric distribution around zero. Therefore,
\begin{align*}
P[(X_{1}^{*}-X_{2}^{*})(Y_{1}^{*}-Y_{3}^{*})>0]=\dfrac{1}{2} P(T)+P[(X_{1}-X_{2})(Y_{1}-Y_{3})>0],
\end{align*}
which is equivalent to
\begin{align*}
 P^{*}(C)=P(C)+\dfrac{1}{2} P(T).
\end{align*}
In the same way, we can show
\begin{equation*}
P^{*}(D)=P(D)+\dfrac{1}{2} P(T).
\end{equation*}
Thence, according to the definition of Spearman's rho in Eq (\ref{eq7}), we can conclude that the continuous extension preserves  Spearman's rho. That is,
\begin{align}\label{eq9}
\rho(X^{*},Y^{*})=\rho(X,Y).
\end{align}

\subsection{Preserving Concordance Order with Continuous Extension }

In this section, we show that the continuous extension of discrete random variables  preserves the concordance order.  This is an important characteristic  that can be used to extend essential properties of the continuous model  to the discrete schemes. Particularly, the preservation of Spearman's rho under the concordance order can be extended  from  random pairs with continuous marginal distributions to  random pairs with  discrete marginal distributions.
 First, we present the definition of concordance order from Yanagimoto and Okamoto (1969).
\begin{definition}\label{def1}
Consider $(X_{1},Y_{1})$ and $(X_{2},Y_{2})$ to be  two random vectors with the same continuous marginal distributions. Then, $(X_{2},Y_{2})$ is more concordant than $(X_{1},Y_{1})$   if
\begin{align}\label{eq10}
P(X_{1}\leq u ,Y_{1}\leq v)\leq P(X_{2}\leq u ,Y_{2}\leq v)
\end{align}
for all $ (u,v)\in  \mathbb{R}^{2}$,  which is denoted by   $(X_{1},Y_{1}) \prec_{c}(X_{2},Y_{2})$.
\end{definition}
If $ X_{1}$  and $Y_{1} $  are  independent, then Eq \eqref{eq10}  can be rewritten as
\begin{align}\label{eq11}
F(u)G(v)\leq P(X_{2}\leq u ,Y_{2}\leq v), ~~~~ \mbox{for all} (u,v)\in  \mathbb{R}^{2},
\end{align}
  where $F(\cdot)$ and $G(\cdot)$ are the distribution functions of $X_1$ and $Y_1$, respectively.  Now, $(X_{1},Y_{1})\prec_{c}(X_{2},Y_{2})$ means that $(X_{2},Y_{2})$ is positively dependent by quadrants (PQD) (see Nelsen, 2006). In other words,  it means that the probability that  $X_{2}$ and $Y_{2}$ to be small is at least as large as it when they are independent.

The definition of concordance ordering given in Definition \ref{def1}   can be extended to the two pairs of $(X_{1},Y_{1})$ and $(X_{2},Y_{3})$ which are used in the definition of Spearman's rho in Eq (\ref{eq5}). Since $ X_{2}$ and $Y_{3}$ in the second pair are independent of each other, therefore the definition of concordance order $(X_{1},Y_{1})\prec_{c}(X_{2},Y_{3})$ in Eq (\ref{eq10}) can be written as follows
\begin{align}\label{eq12}
P(X_{1}\leq u ,Y_{1}\leq v)\leq P(X_{2}\leq u) P(Y_{3}\leq v), ~~~~\mbox{for all} (u,v)\in  \mathbb{R}^{2}.
\end{align}
 This condition  implies that the pair $(X_{1},Y_{1})$ has negative quadrant dependence (NQD).
  Now, assume that for some random pairs $(X_{1},Y_{1})$ and $(X_{2},Y_{3})$ with discrete marginal distributions, the concordance order $(X_{1},Y_{1})\prec_{c}(X_{2},Y_{3})$  defined in Eq \eqref{eq12} holds.  Then, if  $X_{1}(Y_{1})$, $X_{2}(Y_{2})$, and $X_{3}(Y_{3})$ are continued by adding the same  continuous random variable $U (V)$ (see Eq  \eqref{eq8}) such that $U$ and $V$  are independent,  we have
\begin{align*}
P(X^{*}_{1}\leq s ,Y^{*}_{1}\leq t)=&P\left(X_{1}+U\leq s, Y_{1}+V\leq t\right)\\
=&\int_0^1 \int_0^1P\left(X_{1}\leq s-u, Y_{1}\leq t-v\right)h_{U}(u)h_{V}(v)dudv\\
\leq & \int_0^1 \int_0^1 P\left(X_{2}\leq s-u\right) P\left( Y_{3}\leq t-v\right) h_{U}(u)h_{V}(v)dudv \\
=& P(X^{*}_{2}\leq s)P(Y^{*}_{3}\leq t),
\end{align*}
where $h_U(\cdot)$ and $h_V(\cdot)$ are the density functions of $U$ and $V$, respectively.   The second equality follows from Eq \eqref{eq12}.  Therefore,
\begin{align}\label{eq13}
(X_{1},Y_{1})\prec_{c}(X_{2},Y_{3}) \Longrightarrow (X^{*}_{1},Y^{*}_{1})\prec_{c}(X^{*}_{2},Y^{*}_{3}).
\end{align}
Moreover, if $(X,Y)$ is  PQD, then also $(X^{*},Y^{*})$  is PQD. 
Now,  the preservation of Spearman's rho under the concordance order can be concluded from the preservation of concordance order obtained  in Eq \eqref{eq13} and from  the  preservation of Spearman's rho by continuous extension given in Eq  \eqref{eq9}   . That is,
\begin{align*}
(X_{1},Y_{1})\prec_{c} (X_{2},Y_{3})&\Longrightarrow (X^{*}_{1},Y^{*}_{1})\prec_{c} (X^{*}_{2},Y^{*}_{3})\\
&\Longrightarrow^{Yanagimoto} \rho(X^{*}_{1},Y^{*}_{1})\leq \rho(X^{*}_{2},Y^{*}_{3})\\
&\Longleftrightarrow^{from\eqref{eq9}} \rho(X_{1},Y_{1})\leq \rho(X_{2},Y_{3})
\end{align*}
Therefore, when $(X_{1},Y_{1})$, $(X_{2},Y_{2})$ and $(X_{3},Y_{3})$ are three pairs of  discrete random variables, we have
\begin{align}\label{eq14}
(X_{1},Y_{1})\prec_{c} (X_{2},Y_{3})\Longrightarrow \rho(X_{1},Y_{1})\leq \rho(X_{2},Y_{3}).
\end{align}
This means that the  concordance order gives the order of Spearman's rho  in the same direction. Notice that  the inequality between Spearman's rho is strict if  the random pairs $(X_1,Y_1)$ and $(X_2,Y_3)$ are not identically distributed.

 \section{ Copulas and Dependence Measures for Discrete Data}\label{sec3}

In the statistical literature,  it is very common to analyze and investigate associations between bivariate random variables, and then  possibly be extended to deal with multivariate random variables. A copula links   marginal distribution functions  together to construct  a joint distribution function, and completely describes the dependence structure  between the  variables.

Population version of the Kendall's tau and Spearman's rho in terms of copulas and based on concordance and discordance probabilities  for continuous random variables  have been discussed with the details in Joe (1997) and Nelsen (2006). However, in discontinuous cases the probability of tie is not zero, and therefore it needs to be taken into account.  Nikoloulopoulos (2007) proposed  Kendall's tau by using copulas with discrete marginal distributions. More details can be found in Denuit and Lambert (2005), Mesfioui and Tajar (2005) and Nikoloulopoulos (2007).

In this section, we derive and propose  the population version of Spearman's rho via copulas when both random variables are discrete. To this end, let us first introduce  the population version of Kendall's tau  proposed by  Nikoloupolous (2007) for integer-valued discrete random variables  based on concordance and discordance probabilities.
Let $X$ and $Y$ be  discrete random variables taking integer values. Moreover, assume  $H(\cdot, \cdot)$ and $h(\cdot, \cdot)$ are the joint distribution function  and joint mass function, respectively, in which  $F(\cdot)$ and $G(\cdot)$ are the marginal distributions of $X$ and $Y$, respectively, with mass functions $f(\cdot)$ and $g(\cdot)$. Then, the population version of Kendall's tau of discrete random variables $X$ and $Y$  with copula $\mathcal{C}(\cdot, \cdot)$ is obtained  as

\begin{align}\label{eq3.1}
\tau (X,Y)=\sum _{x=0}^{\infty }\sum _{y=0}^{\infty }h(x,y)  \left\{ 4 \mathcal{C}(F(x-1),G(y-1))-h(x,y)\right\} +\sum _{x=0}^{\infty }\left(f^{2} (x)+g^{2} (x)\right) -1,
\end{align}
where
\begin{align}\label{eq3.2}
h(x,y)=\mathcal{C}(F(x),G(y))-\mathcal{C}(F(x-1),G(y))-\mathcal{C}(F(x),G(y-1))+\mathcal{C}\left(F(x-1),G(y-1)\right)
\end{align}
is the joint pmf of $X$ and $Y$, $\tau(X, Y)$ is the  Kendall's tau  of  $X$ and $Y$.

Now, similar to   Nikoloupolous (2007), we formulate and derive the  population version of Spearman's rho of discrete random variables as follows.

\begin{theorem}\label{thr1}
Assume $X$ and $Y$ are  integer-valued discrete random variables with the joint distribution function  $H(\cdot, \cdot)$ and the joint mass function $h(\cdot, \cdot)$, in which  $F(\cdot)$ and $G(\cdot)$ are the marginal distribution functions $X$ and $Y$, respectively, with mass functions $f(\cdot)$ and $g(\cdot)$.  The population version of Spearman's rho of $X$ and $Y$, $\rho^S(X,Y)$, with copula $\mathcal{C}(\cdot, \cdot)$ is  obtained as
\begin{align}\label{eq3.4}
\begin{split}
\rho^S (X,Y)=&6 \sum _{x=0}^{\infty }\sum _{y=0}^{\infty }h(x,y)  \left[ (1-F(x))(1-G(y))+F(x-1)G(y-1)-\dfrac{1}{2}f(x)g(y)\right] \\
&~~~~~~~~~~~~~~~~~~~+3\sum _{x=0}^{\infty } \left(f^{2} (x)+g^{2} (x)\right) -3,
\end{split}
\end{align}
where  $h(x,y)$ is the joint  pmf of $X$ and $Y$ defined in Eq \eqref{eq3.2}.
 \end{theorem}
The proof is provided in the Appendix.

\subsection{Spearman's Rho of Bernoulli Random Variables}

 Since the Bernoulli random variable takes only two values zero and one, it is easy to derive the closed form expression for  Spearman's rho  of two Bernoulli random variables $X$ and $Y$  by using Eq \eqref{eq3.4}. 

\begin{theorem}\label{thr2}
Let $X$ and $ Y$ be  two  Bernoulli random variables with success probabilities of  $p_{X}$ and $p_{Y}$, respectively. Then, the Spearman's rho correlation between $X$ and $Y$ based on the  copula $\mathcal{C}(u, v)$  is  
\begin{align}\label{eq3.5}
\rho^S(X,Y)=-3+3\mathcal{C}(1-p_{X} ,1-p_{Y})+3p_{X}+3p_{Y}-3p_{X}p_{Y}.
\end{align}
\end{theorem}
The proof is given in the Appendix. For comparison of  the Spearman's rho  and Kendall's tau  in this case, notice that  Nikoloupolous (2007) derived the Kendall's tau of binary random variables as
\begin{equation}\label{eq3.6}
  \tau(X,Y)= 2\left[\mathcal{C}(1-p_X, 1-p_Y) - (1-p_X)(1-p_Y)  \right].   
\end{equation}

\subsection{Upper and Lower Bounds of Spearman's rho for Binary Margins }
Using the Fr\'{e}chet-Hoeffding bounds for copulas, Nikoloupolous  (2007) showed that  the lower and upper bounds of  Kendall's tau for binary random variables  are  $-0.5$ and $0.5$, respectively.  Similarly, we used the  Fr\'{e}chet-Hoeffding  bounds and Eq \eqref{eq3.4} to obtain the lower and upper bounds of Spearman's rho of binary random variables.  More details of Fr\'{e}chet-Hoeffding bounds  can be found in  Nelsen (2006), Joe (2014),  and Hofert et al. (2018).
\begin{theorem}\label{thr3}
Using the Fr\'{e}chet-Hoeffding bounds, it can be shown that the lower and upper bounds of Spearman's rho  for   binary  random variables  are  $-0.75$ and $0.75$, respectively.
\end{theorem}
{\bf Proof:}
The proof follows from the linear relationship $\rho^{S}(X,Y)=1.5\,\tau(X,Y)$  derived  from  Eqs (19)  and (20),  and  using the  lower and upper bounds of Kendall’s tau with Bernoulli margins  proposed   by Nikoloupolous (2007).
It can be shown that  $\rho^S(X,Y)$ reaches its maximum and minimum values  when $p_{X}=p_{Y}= 0.5$, that is, $-  0.75\leq\rho^S(X,Y)\leq 0.75.$ $\blacksquare$

\section{Simulation Study}\label{sec4}

In this section, we  conduct Monte Carlo simulation studies to investigate the behavior of the proposed Spearman's rho correlation of discrete variables with some specific  discrete marginal distributions. Moreover, several well-known Archimedean copula families such as the Frank, Gumbel and Clayton copulas  are used in the numerical analysis. In addition, the results of the Spearman's rho correlation of count data are compared with their corresponding Kendall's tau values.  The population version of Kendall's tau of count data is proposed by Nikoloupolous and Karlis (2009).

For the purpose of comparison,   Spearman's rho and  Kendall's tau  are calculated with  different  marginal distributions, i.e.,  Poisson, Bernoulli, and Negative Binomial distributions.

Five different  copula functions are presented in Table \ref{Table5}. In  Table \ref{Table5}, $\theta$ denotes the dependence parameter and shows the strength of dependency between two  random variables.  For instance, in the Frank copula,  as $\theta$ goes to zero  it represents independence, whereas as  $\theta$ goes to infinity, it describes perfect dependence.   See for example  Nelsen (2006),  Joe (2014),  and Hofert et al. (2018)  for more details about the copula families  provided in Table \ref{Table5}.  Once we estimate the copula dependence parameter, we can calculate Spearman's rho and Kendall's tau values by using Eqs  \eqref{eq3.2} and \eqref{eq3.1}, respectively.

\begin{table}[h]\caption{Archimedean copulas and their corresponding generating functions  $\phi(t)$   }\label{Table5}
{\small
\begin{center}
\begin{tabular}{l c c c}
\hline
Family                &                      $\phi(t)$                            &  $\theta \in$      &$ \mathcal{C}(u_1,u_2;\theta)$ \\
\hline
Frank                 &-$\ln \dfrac{e^{-\theta t}-1}{e^{-\theta}-1}$  &  $\theta \neq0$ &$ -\frac{1}{\theta}\ln\bigl[1+\dfrac{(e^{-\theta u_1}-1)(e^{-\theta u_2}-1)}{e^{-\theta}-1}\bigr] $ \\
Clayton               &   $t^{-\theta}-1 $                                    &   $\theta>0$      &$ (u_1^{-\theta}+u_2^{-\theta}-1)^{-\frac{1}{\theta}}$ \\
Gumbel-Hugard   &              $(-\ln t)^\theta $                       &  $\theta\geq 1$   &$\exp \Bigl\lbrace -\bigl[(-\ln u_1)^\theta+(-\ln u_2)^\theta\bigr]^{\frac{1}{\theta}}\Bigr\rbrace$\\
Ali-Mikhail-Haq     &   $ \ln \dfrac{1-\theta(1-t)}{t} $                &$-1\leq \theta <1$&$\dfrac{u_1 u_2}{1-\theta(1-u_1)(1-u_2)}$ \\
Joe                     &   $ -\ln(1-(1-t)^{\theta}) $                &$\theta\geq 1$      &$1-\bigr[ (1-u_1)^{\theta} + (1-u_2)^{\theta} -(1-u_1)^{\theta}(1-u_2)^{\theta} \bigr]^{1/ \theta}$ \\
\hline
\end{tabular}
\end{center}}
\end{table}

\begin{figure}[ht!]
\centering
\includegraphics[width=14cm,height=9.8cm]{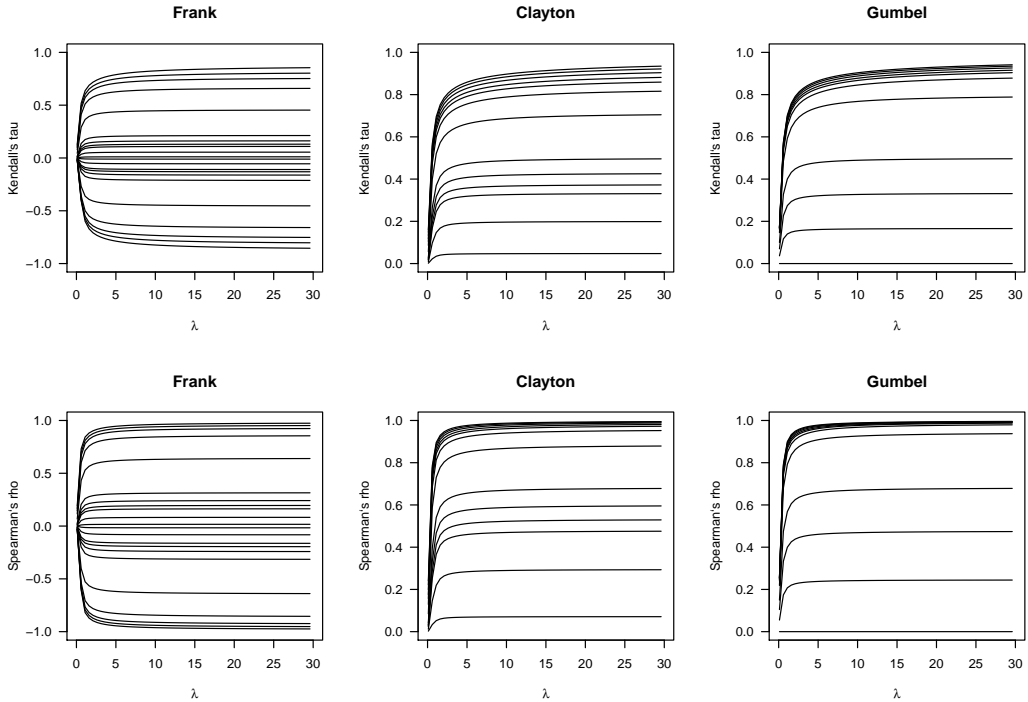}
\caption{Kendall's tau and Spearman's rho values computed using Frank, Clayton, and Gumbel copulas, and Poisson marginal distributions with the same parameter $\lambda$ from one to 30. Larger value of the copula parameter lead to a higher curve.}
\label{f1}
\end{figure}

\begin{figure}[ht!]
\centering
 \includegraphics[width=14cm,height=9.8cm]{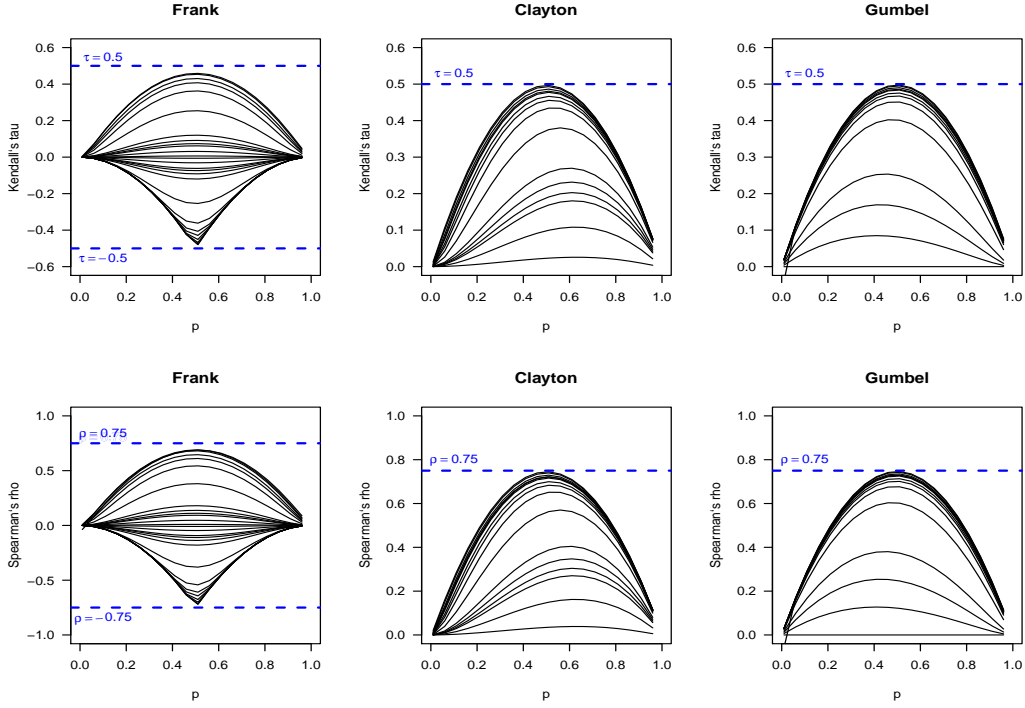}
\caption{Kendall's tau versus Spearman's rho values computed using Frank, Clayton and Gumbel copula and Bernoulli marginal distributions with the same parameter.}
\label{f2}
\end{figure}

 Figure \ref{f1} shows the comparison of   the Spearman's rho and  Kendall's tau  values obtained from Poisson marginal distributions with different values of  the parameter $\lambda$.  Each curve in the figure corresponds to a  different value of the  copula parameter $\theta$, 
  where  higher curves correspond to higher values of the copula parameter $\theta$.  Similarly, Figure \ref{f2} displays the comparison of  the Spearman's rho and  Kendall's tau computed from Bernoulli marginal distributions with parameter $p$, $0<p<1$. As in Figure \ref{f1},  the top row in Figure \ref{f2} shows the Kendall's tau obtained under three different copula functions, and the bottom row shows the Spearman's rho computed under the Frank, Clayton, and Gumbel copulas.

Note that  the Frank copula function is the only symmetric copula here that permits both negative and positive dependence, whereas the Gumbel and Clayton copulas are only able to capture positive dependence.  These properties of copula functions  can be seen in  Figures \ref{f1} and \ref{f2}. Furthermore, both of the  Spearman's rho and  Kendall's tau are increasing functions of the copula parameter $\theta$.

Moreover, since the Frank copula is flexible and can capture both positive and negative associations, in our simulation study, we consider both positive and negative values of the copula parameter $\theta$ for the Frank copula, whereas, only positive values of $\theta$ are used for Gumbel and Clayton copulas.

Similarly,  Spearman's rho and  Kendall's tau  are computed  based on the same copula functions  but with Bernoulli marginal distributions. Recall that, in Theorem \ref{thr3}, we showed that the upper and lower bounds of Spearman's rho in this case are $0.75$ and $-0.75$, respectively. However, Nikoloupolous and Karlis (2009) showed that the upper and lower bounds of Kendall's tau for Bernoulli random variables are $0.5$ and $-0.5$, respectively. Figure \ref{f2} displays the corresponding  Spearman's rho and  Kendall's tau values calculated from Bernoulli marginal distributions with the same parameter $p$.

Table \ref{table1} reports the Monte Carlo simulation results when data are generated from  Frank, Gumbel and Clayton copulas with the discrete margins following a  Negative Binomial ($NB(r,p)$) distribution  that counts  the number of failures until $r$ successes  with $r=3$ and $p=0.4$. Three different values of the copula parameter   are selected in order to obtain the Spearman's rho and  Kendall's tau correlations, i.e, $\theta=0.5, 1, 3$ for Frank and Clayton, and $\theta=1.5, 2, 3$ for Gumbel. For each copula parameter, we consider sample sizes of $n=100, 300$, and $800$. The copulas are estimated by using the log-likelihood function of the function proposed in Eq \eqref{eq3.2}. One thousand iterations are performed, and the mean and standard deviation of the estimators are obtained.  The parameter estimates for $\tau$ and $\rho$ reported in Tables \ref{table1}-\ref{table3} are plug-in estimates obtained from their explicit expression given in Eqs \eqref{eq3.1} and \eqref{eq3.4}.  The results of Table \ref{table1} show that the maximum likelihood estimators (MLEs) are consistent,  that is, when sample size increases, the estimated parameters converge to their true values.  In order to  better understand  the relationship between  these two  measures of dependence,  the estimated ratio of  Spearman's rho to  Kendall's tau for each case  is provided in the last column of Tables \ref{table1}-\ref{table3}.  The results show that the  ratio of  Spearman's rho to  Kendall's tau  is always greater than one, and the maximum ratio reaches to  $1.5$.    


\begin{table}[!htbp]
\caption{Simulation results with Negative Binomial margins with $r=3$ and  $p=0.4$ 
}\label{table1}
 \centering
 {\footnotesize
\begin{center}
\begin{tabular}{c|c c c c c c c c}

Family & $\theta$&  $\tau$& $\rho$ & $n$ & $\hat{\theta}$ (sd)  &   $\hat{\tau}$ (sd)  & $\hat{\rho}$ (sd)  & $\hat{\rho}/\hat{\tau}$\\
\hline

         &   $0.5$   &  0.054  & 0.081  & 100  &   0.484 (0.621)       &  0.021 (0.067)         &   0.031 (0.100)  &   1.476 \\
         &              &            &          & 300   &  0.498 (0.352)        &  0.043 (0.038)        &  0.064 (0.057)    &  1.488 \\
 \vspace{0.15cm}
         &               &            &          &  800 &   0.500 (0.213)       &  0.050 (0.023)         &  0.075 (0.034)    &1.500   \\
          & $1$         &  0.108  & 0.161 &  100 &  1.023 (0.632)         &   0.075 (0.068)      &  0.112 (0.102)     &  1.493 \\
Frank  
          &               &            &          & 300  &  0.986 (0.371)        &   0.094 (0.040)       &  0.140 (0.059)   &   1.489 \\
  \vspace{0.15cm}
          &                &            &          & 800  &  0.998 (0.215)       &  0.103  (0.022)       &  0.154  (0.033)    & 1.495  \\
           & $3$         & 0.300    & 0.439 &  100 &  3.046 (0.686)       &  0.258 (0.061)         &  0.374 (0.086)   &  1.450 \\
           &               &             &           & 300 & 3.015 (0.383)         &      0.285 (0.034)    &       0.416 (0.047)  &1.460  \\
  \vspace{0.15cm}
           &                &             &          & 800 &  3.003 (0.234)       &    0.294 (0.020)        &   0.430 (0.028)   &  1.463 \\
   &       20      &       0.773   &       0.937   &       100     &       20.485 (2.454) &       0.722   (0.045) &       0.858   (0.064) &       1.189   \\
   &           &         &         &       300     &       19.899
(1.329) &       0.752   (0.018) &       0.906   (0.022) &       1.205\\
  &             &          &          &       800     &       20.107
 (0.867) &       0.767   (0.007) &       0.927   (0.008) &       1.208\\

\hline
           & $0.5$       & 0.193     & 0.286  & 100 &  0.345 (0.282)       &  0.155 (0.061)          &  0.228 (0.088)   & 1.471 \\
           &                &              &          & 300  &  0.505 (0.101)       &  0.182 (0.032)         &  0.268 (0.046)   &  1.473 \\
\vspace{0.15cm}
           &                &              &           & 800  &  0.502 (0.061)       &   0.189  (0.007)       &  0.279 (0.011)  & 1.476 \\
           & $1$          &  0.321     &  0.464 & 100  & 1.022 (0.225)        &  0.282 (0.052)         &  0.404 (0.072)  &  1.433 \\
Clayton 
           &                &               &          & 300   &  1.007 (0128)        &  0.308 (0.028)         &  0.444 (0.039)  & 1.442 \\
 \vspace{0.15cm}
           &                &               &           & 800  &  1.002 (0.077)       &   0.316 (0.017)       &   0.457 (0.022)   & 1.446 \\
           & $3$           &  0.572     & 0.766  & 100   &  3.048 (0.446)      &   0.523 (0.047)       &   0.691 (0.061)  & 1.321 \\
           &                 &               &          & 300   &  3.020 (0.249)       &   0.556 (0.022)       &   0.742 (0.026)  & 1.335 \\
  \vspace{0.15cm}
           &                 &                &         & 800   &  2.994 (0.152)       &   0.565 (0.012)       &    0.756 (0.014) & 1.338 \\
&       20      &       0.837   &       0.968   &       100     &       20.878  (3.638) &       0.784   (0.036) &       0.887   (0.052) &       1.132  \\
&           &        &       &       300     &       20.393 (1.675) &       0.819   (0.012) &       0.939   (0.017) &       1.147 \\
 &        &      &     &       800     &       20.152 (1.219) &    0.829   (0.007) &       0.956   (0.008) &   1.152  \\

\hline
           &   $1.5$       &  0.326      & 0.467 & 100  &   1.515 (0.130)       &  0.280 (0.067)       &  0.396 (0.093)  & 1.414\\
           &                  &                &          & 300  &  1.506 (0.069)      &    0.311 (0.032)   &  0.444 (0.043)  & 1.427\\
  \vspace{0.15cm}
           &                  &                &          & 800  &  1.501 (0.043)     &   0.320  (0.019)   &   0.458  (0.026)  &1.431 \\
           &  $2$           &   0.487      & 0.668 & 100  &  2.011 (0.167)        &   0.440 (0.057)      &   0.597 (0.076)  &1.357 \\
Gumbel  
           &                  &                 &          & 300  &  2.010 (0.098)  &   0.472 (0.028)  &   0.646 (0.035)  &1.369 \\
 \vspace{0.15cm}
           &                   &                &          & 800  &  2.001 (0.060)  &   0.481 (0.015)       &   0.660 (0.018) & 1.372 \\
           & $3$            & 0.644         & 0.831 & 100  &  3.025 (0.275)        &   0.601 (0.049)       &   0.766 (0.065)   & 1.275 \\
           &                  &             &          & 300 &  3.003 (0.159)         &   0.629 (0.021)       &  0.808 (0.025)  & 1.285 \\
  \vspace{0.15cm}
           &                  &           &          & 800 &  3.001 (0.093)        &   0.639  (0.011)      &   0.822  (0.012)  & 1.286\\
 &       20      &       0.869   &       0.977   &       100     &       22.031 (8.270)  &       0.841   (0.030)  &       0.934   (0.042) &       1.111  \\
 &            &       &        &       300     &       20.788
(2.922) &       0.858   (0.012) &       0.960    (0.017) &    1.118  \\
&         &       &    &       800     &       19.983  (1.510)  &       0.864 (0.005) &       0.970    (0.006) &       1.122 \\
\end{tabular}
\end{center}}
\end{table}

Comparing the performance of the copula-based Spearman's rho and  Kendall's tau with discrete margins shows that Spearman's rho takes a wider range of values than does Kendall's tau.   This  is because of a functional relationship between these two measures of dependence, e.g.,  there is a simple  linear relationship  $\rho^{S}(X,Y)=1.5\,\tau(X,Y)$ when the marginal distributions are  Bernoulli (see  Theorem 3.2 and Theorem 3.3).   When the marginal distributions are not Bernoulli, this relationship is not linear but a function of the copula parameter and the parameter of the marginal distributions.   Figure 3 shows the functional  relationship  between  these two  measures  with different marginal distributions and different values of  the copula parameter obtained under three different copula functions.



\begin{table}[!htbp]
\caption{Simulation results with  Poisson margins with $\lambda=0.5$ 
}\label{table2}
\begin{center}
\scalebox{0.8}{%
\begin{tabular}{c|c c c c c c c c }
    Family &   $\theta$ &   $\tau$&   $\rho$   &    $n$   &   $\hat{\theta}$ (sd)   &   $\hat{\tau}$ (sd)   &   $\hat{\rho}$ (sd) & $\hat{\rho}/\hat{\tau}$\\
\hline
    &    $ 0.5$ &  0.031 & 0.047 &  $100$ &   0.518 (0.840)    &  0.021 (0.049)     &    0.031 (0.074) &1.476\\
     &   & &  & 300     &       0.376 (0.452) &       0.021   (0.028) &       0.031   (0.042) &       1.499 \\
 \vspace{0.15cm}
      &   & & &    $800$  &  0.503 (0.287)     &   0.030  (0.018)    &    0.045 (0.027) &1.500\\
     &  $ 1$&   0.062&  0.094  &   $100$  &  1.021 (0.874)   &   0.050 (0.051)   &   0.075 (0.077) &1.500\\
   Frank  
     &   && &  $300$  &  0.965 (0.477)        &    0.057 (0.029)       &   0.085 (0.044) &1.491\\
 \vspace{0.15cm}
     &   & & & $800$  &  1.027 (0.284) &        0.062 (0.017) &         0.093 (0.026) &1.500\\
    &  $ 3$&  0.176 & 0.263 &   $100$  &  3.076 (0.989)     &   0.158 (0.049)     &   0.236 (0.072) &1.494\\
     &   &&&   $300$  &  3.053 (0.539) &         0.172 (0.027)       &   0.257 (0.040) &1.494\\
 \vspace{0.15cm}
      &   && &  $800$ &  3.001 (0.344)       &   0.173  (0.017)      &    0.259 (0.026) &1.497\\
      &       20      &       0.448   &       0.648   &       100     &       20.716 (5.617) &       0.416   (0.034) &       0.601   (0.049) &       1.444 \\
      &     &     &     &       300     &       21.031 (3.467) &       0.443   (0.013) &       0.64    (0.016) &       1.445\\
      &      &     &       &       800     &       20.804 (2.269) &       0.446   (0.009) &       0.645   (0.012) &       1.446 \\
\hline
    &  $ 1$&  0.143  & 0.214 &  $100$ &   1.061 (0.473)      &   0.128 (0.048)     &    0.192 (0.072) &1.500\\
     &   & &&  $300$  &  1.016 (0.271) &          0.138 (0.029)      &   0.207 (0.043) &1.500\\
 \vspace{0.15cm}
      &   &&  & $800$  &  0.999 (0.163)       &   0.140  (0.017)       &    0.210 (0.026) &1.500\\
     &  $ 2$&  0.228 &  0.342 &   $100$  &  2.113 (0.713)     &    0.211 (0.047)     &   0.315 (0.070) &1.493\\
   Clayton  
    &   & &&  $300$  &  2.029 (0.383)       &   0.223 (0.027)      &   0.333 (0.039) &1.493\\
 \vspace{0.15cm}
      &   &&&   $800$ &  2.030 (0.226)     &    0.227 (0.015)     &    0.339 (0.023) &1.493\\
    &   $ 3$&  0.285& 0.426  &   $100$  &  3.159 (0.932)    &   0.264 (0.043)     &    0.394 (0.063) &1.492\\
     &  & &&  $300$ &  3.030 (0.478)       &   0.278 (0.023)       &   0.415 (0.034) &1.493\\
 \vspace{0.15cm}
      &   & & & $800$  &  3.017 (0.319)      &   0.283  (0.015)      &   0.422  (0.022) &1.491\\
      & 20       &  0.475    &  0.685    &  100      & 24.058   (5.041) &        0.449    (0.030)   &        0.646    (0.042)  &  1.437 \\
 &           &      &     &       300     &       20.439  (3.920)  &       0.466   (0.012) &       0.671   (0.015) &       1.441 \\
  &        &        &          &       800     &       20.049 (2.324) &       0.470   (0.007) &       0.678   (0.009) &       1.442 \\
\hline
    &  $ 1.5$&0.209 & 0.309  &  $100$ &   1.526 (0.167)    &   0.185 (0.046)      &    0.272 (0.067) &1.470\\
    &   & & & $300$  &  1.508 (0.094)      &    0.202 (0.025)       &   0.298 (0.036) &1.475\\
 \vspace{0.15cm}
      &   && &   $800$  &  1.502 (0.057)     &    0.206 (0.015) &        0.304  (0.022) &1.476\\
    &  $ 2$&  0.302&  0.440 &   $100$  &  2.061 (0.028)     &   0.277 (0.045)     &   0.403 (0.064) &1.455\\
  Gumbel  
    &   &&&   $300$  &  2.018 (0.152)      &    0.295 (0.021)       &   0.430 (0.029) &1.458\\
 \vspace{0.15cm}
      &   & &&  $800$  &  2.008 (0.094)     &   0.299  (0.013)      &    0.436 (0.018) &1.458\\
    &  $ 3$& 0.385& 0.554 &   $100$  &  3.162 (0.728)      &    0.362 (0.038)      &   0.520 (0.054)  &1.436\\
    &   & &&  $300$  &  3.025 (0.330) &         0.378 (0.019) &         0.543 (0.026) &1.437\\
 \vspace{0.15cm}
      &   &  && $800$  &  3.021 (0.195)       &    0382 (0.011) &         0.550 (0.015) &1.440\\
      &        20       &        0.513    &        0.721    &    100      &   22.984   (6.399) &        0.495    (0.029)  &     0.694    (0.041)  &       1.402   \\
&          &       &    &       300     &       21.228
(3.566) &       0.507   (0.009) &       0.713   (0.013) &     1.405  \\
& &       &    &       800     &       20.936  (2.243) &       0.510   (0.006) &       0.717   (0.008) &    1.406  \\
\end{tabular}}
\end{center}
\end{table}

Similarly, Table \ref{table2} reports the simulation results when data are generated from the same copula functions but with the same margins following  Poisson  distributions with $\lambda=0.5$. However, Table \ref{table3} shows  the Monte Carlo simulation  results when data are generated by the Frank, Gumbel and Clayton copulas but with different marginal distributions, one margin is Negative Binomial  with $r=3$ and $p=0.4$, and the other is Poisson with $\lambda= 0.4$.

\begin{table}[!htbp]
\caption{Simulation results with two different margins: $Poisson (\lambda=0.5$) and  $NB(r=3, p=0.4$) 
}\label{table3}
\begin{center}
\scalebox{0.8}{%
\begin{tabular}{c|c c c c c c c c }
Family & $\theta$& $\tau$& $\rho$  &  $n$ & $\hat{\theta}$(sd)  & $\hat{\tau}$ (sd) &  $\hat{\rho}$ (sd)& $\hat{\rho}/\hat{\tau}$\\
\hline
 & $0.5$&0.041&0.061& $100$& 0.497 (0.697) & 0.022 (0.055)  & 0.033 (0.082)  & 1.500\\
&&&& $300$ &0.499 (0.400) & 0.035 (0.032)  & 0.052 (0.048) & 1.486\\
 \vspace{0.15cm}
&&&& $800$ &0.493 (0.244) & 0.037 (0.020)  & 0.055 (0.030)  & 1.486\\
 & $2$& 0.157& 0.235& $100$& 2.024 (0.769) & 0.132 (0.054) & 0.197 (0.081)  & 1.492\\
Frank 
&&&& $300$& 2.005 (0.427)   & 0.148 (0.030)   & 0.222 (0.045) & 1.500\\
 \vspace{0.15cm}
&&&& $800$ &1.996 (0.265) & 0.153 (0.019)  & 0.229 (0.028)  & 1.497\\
 &$3$& 0.224 &0.333& $100$& 3.071 (0.854) & 0.196 (0.054) & 0.291 (0.079)  & 1.485 \\
& &&&$300$& 3.027 (0.480)  &  0.215 (0.029)  & 0.320 (0.043) & 1.488\\
 \vspace{0.15cm}
&&& &$800$ &3.014 (0.282)  & 0.217 (0.017)  & 0.321 (0.025)  & 1.479\\
&       20      &       0.5     &       0.714   &       100     &       20.730 (4.208) &       0.453 (0.036) &       0.643 (0.051) &       1.419\\
&&& &      300     &       20.076 (2.471) &       0.486   (0.013) &       0.693   (0.019) &       1.424 \\
&&& &      800     &       20.233 (1.316) &       0.494   (0.006) &       0.704   (0.008) &       1.425  \\
\hline
 & $1$&  0.203&0.304& $100$& 1.039 (0.333) &  0.178 (0.046)  &  0.265 (0.068) & 1.489 \\
&&  && $300$ &1.020 (0.199) &   0.196 (0.026)  &  0.293 (0.039)   & 1.495 \\
 \vspace{0.15cm}
&&&& $800$ &1.005 (0.120)  &    0.200 (0.016) &     0.299 (0.024)  & 1.495 \\
 & $2$&  0.303&  0.451& $100$& 2.098 (0.513) &  0.275 (0.040)  &  0.409 (0.059)  &1.487 \\
Clayton 
&&&& $300$ &2.008 (0.294) &    0.293 (0.022) &    0.435 (0.033)   & 1.485\\
 \vspace{0.15cm}
&&& &$800$ &2.010 (0.177) &    0.299 (0.013) &    0.448 (0.019) & 1.498\\
 & $3$&  0.362&  0.536& $100$& 3.084 (0.720)  &  0.328 (0.039) &  0.484 (0.057)  & 1.476\\
&& &&$300$& 3.028 (0.416) &    0.351 (0.021)   &  0.520 (0.030)  & 1.481 \\
 \vspace{0.15cm}
&&&& $800$ &3.020 (0.246)   &  0.356 (0.012)   &  0.525 (0.017) & 1.475 \\
 &       20      & 0.513   &0.729   &100     &       23.145 (7.749) &   0.478   (0.033) &       0.677   (0.048) &       1.415 \\
&&&&       300     &       20.954 (3.761) &       0.499   (0.013) &       0.708   (0.019) &       1.418  \\
&&&&        800     &       20.502 (2.126) &       0.508   (0.006) &       0.721   (0.009) &       1.420\\
\hline
 &$1.5$&0.253&0.372 & $100$& 1.516 (0.148) &0.217 (0.052) &0.317 (0.075)  & 1.461 \\
&&&& $300$ &1.503 (0.086) &0.240 (0.028) &0.351 (0.040)  & 1.463\\
 \vspace{0.15cm}
&&&& $800$& 1.501 (0.051) &0.249 (0.016) &0.364 (0.023)  & 1.462\\
 & $2$&0.363&0.524& $100$& 2.036 (0.230)  &0.329 (0.045) &0.473 (0.064)  &1.438 \\
Gumbel 
&&&& $300$& 2.011 (0.130)  &0.351 (0.023) &0.505 (0.032) & 1.439\\
 \vspace{0.15cm}
&&&& $800$ &2.009 (0.083) &0.359 (0.013)  &0.518 (0.018) & 1.443\\
 &$3$ &0.452&0.643& $100$&3.119 (0.474) &0.420 (0.039) &0.594 (0.056)  & 1.414\\
&&& &$300$& 3.028 (0.246) &0.441 (0.016) &0.626 (0.022) & 1.420\\
 \vspace{0.15cm}
&&&& $800$ &3.012 (0.147) &0.448 (0.009) &0.636 (0.012)  & 1.420\\
& 20  &  0.528    &        0.741    &        100      & 24.719   (7.368) &   0.502    (0.024)  &        0.701    (0.036)  &  1.398 \\
&&&&300      & 21.705   (3.114) & 0.518    (0.011)  &  0.726    (0.017)  & 1.401\\
&&&& 800     &  20.849 (2.900)   &  0.524   (0.005) &       0.735   (0.007) &       1.402 \\
\end{tabular}}
\end{center}
\end{table}

Figure 3 displays the  ratio of  Spearman's rho to  Kendall's tau  versus the parameter of the marginal distributions where each curve represents a different  value of  the copula parameter.   The top row is obtained      with $Bin(5,p)$ marginal distributions, the middle row is computed with $NB(4, p)$ marginals, and the bottom row is obtained with $Poisson(\lambda)$ marginals, each with  three  copula functions.      The plots reveal that the relationship between  Spearman's rho and   Kendall's tau  is  not linear but when the marginals are Binomial  it tends to follow a U-curve pattern. For the two other cases,   the relationship is not linear but tends to a convex pattern.  The maximum ratio of  Spearman's rho to  Kendall's tau  reaches to  $1.5$ .

 \begin{figure}[!htbp]
   \centering
 \begin{subfigure}[b]{0.34\textwidth}
                \includegraphics[width=\textwidth]{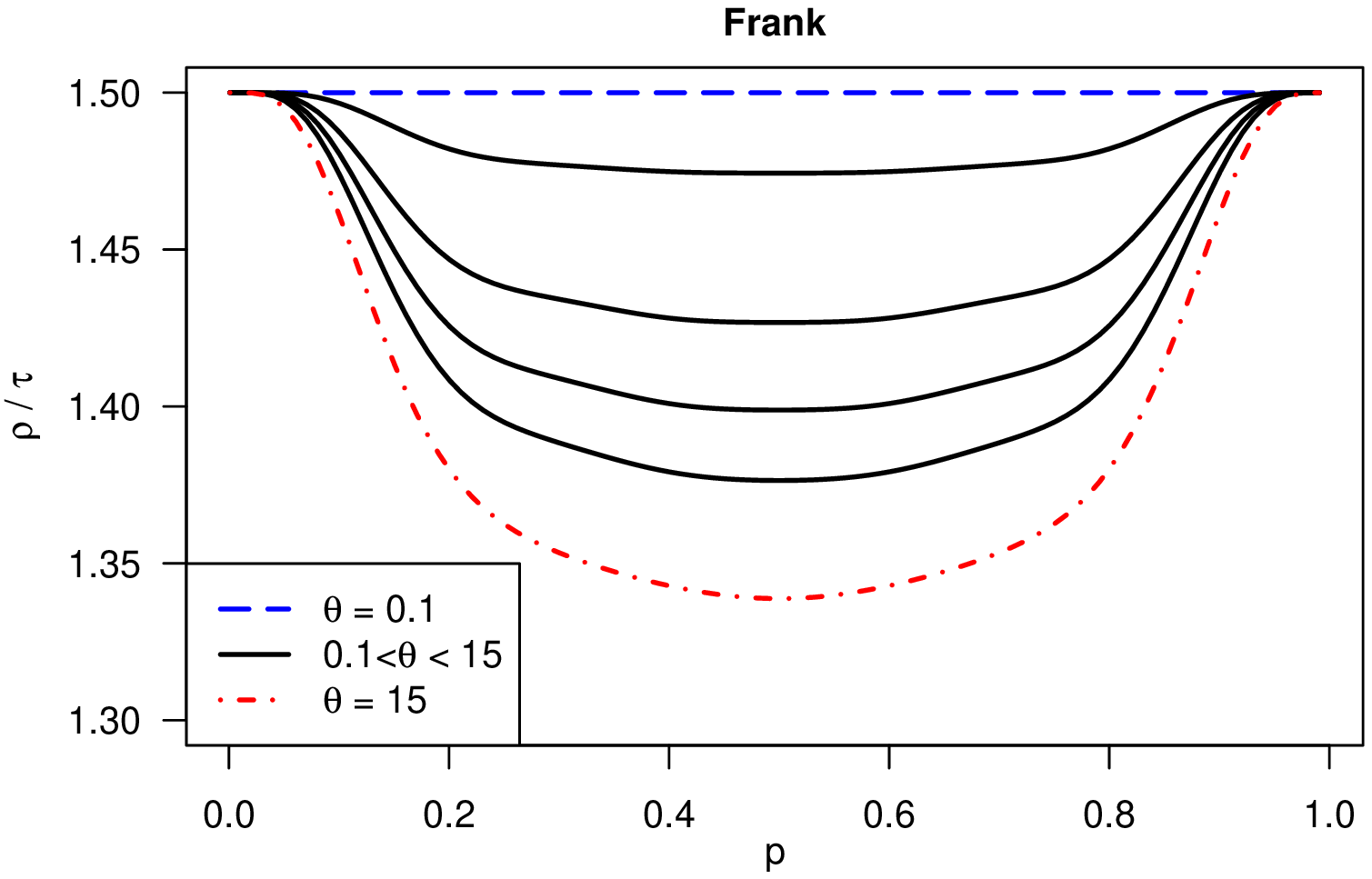}
        \end{subfigure}%
         \hspace{-.4cm}
           \begin{subfigure}[b]{0.34\textwidth}
                \includegraphics[width=\textwidth]{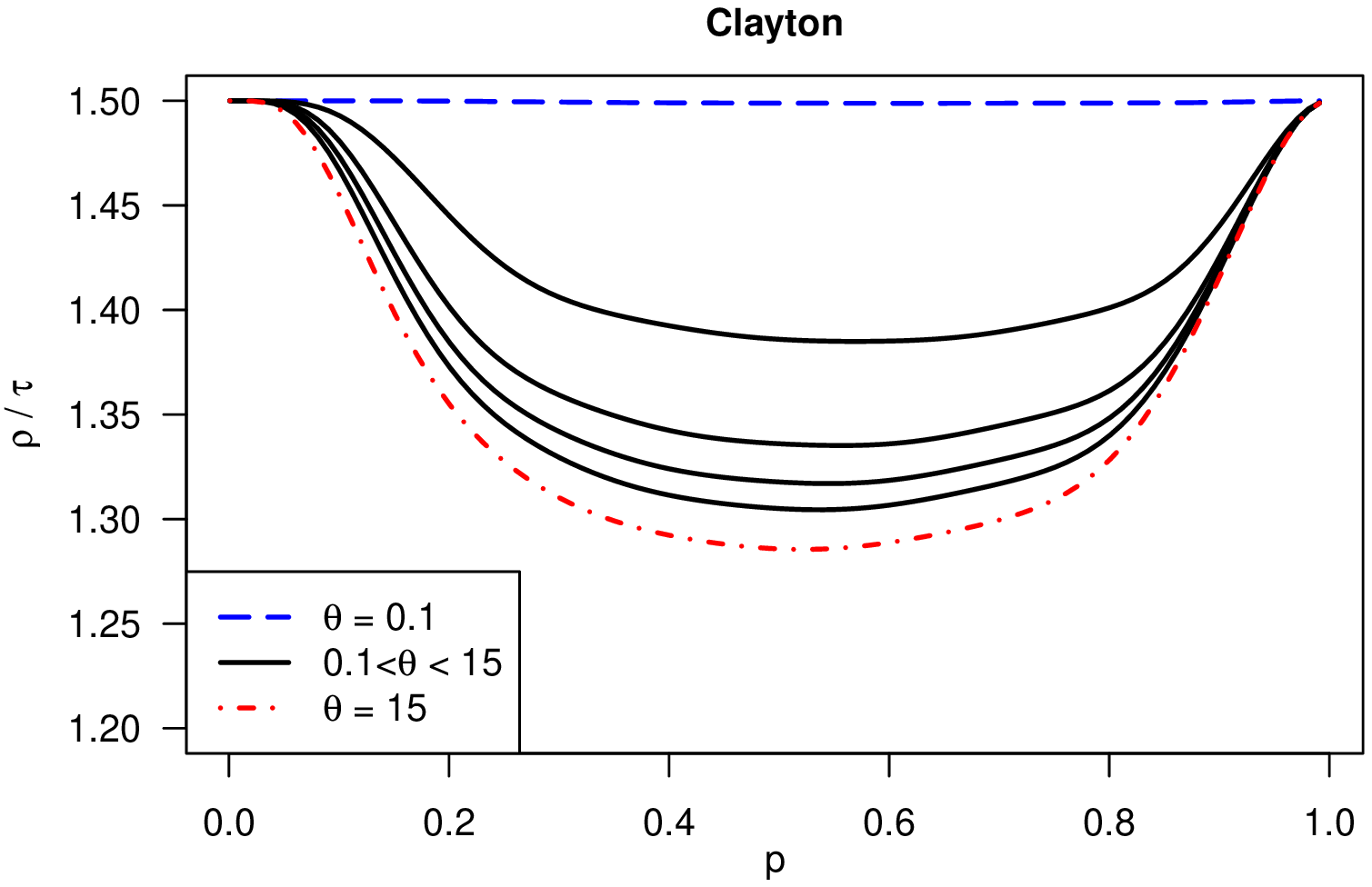}
        \end{subfigure}
          \hspace{-.4cm}
         \begin{subfigure}[b]{0.34\textwidth}
                \includegraphics[width=\textwidth]{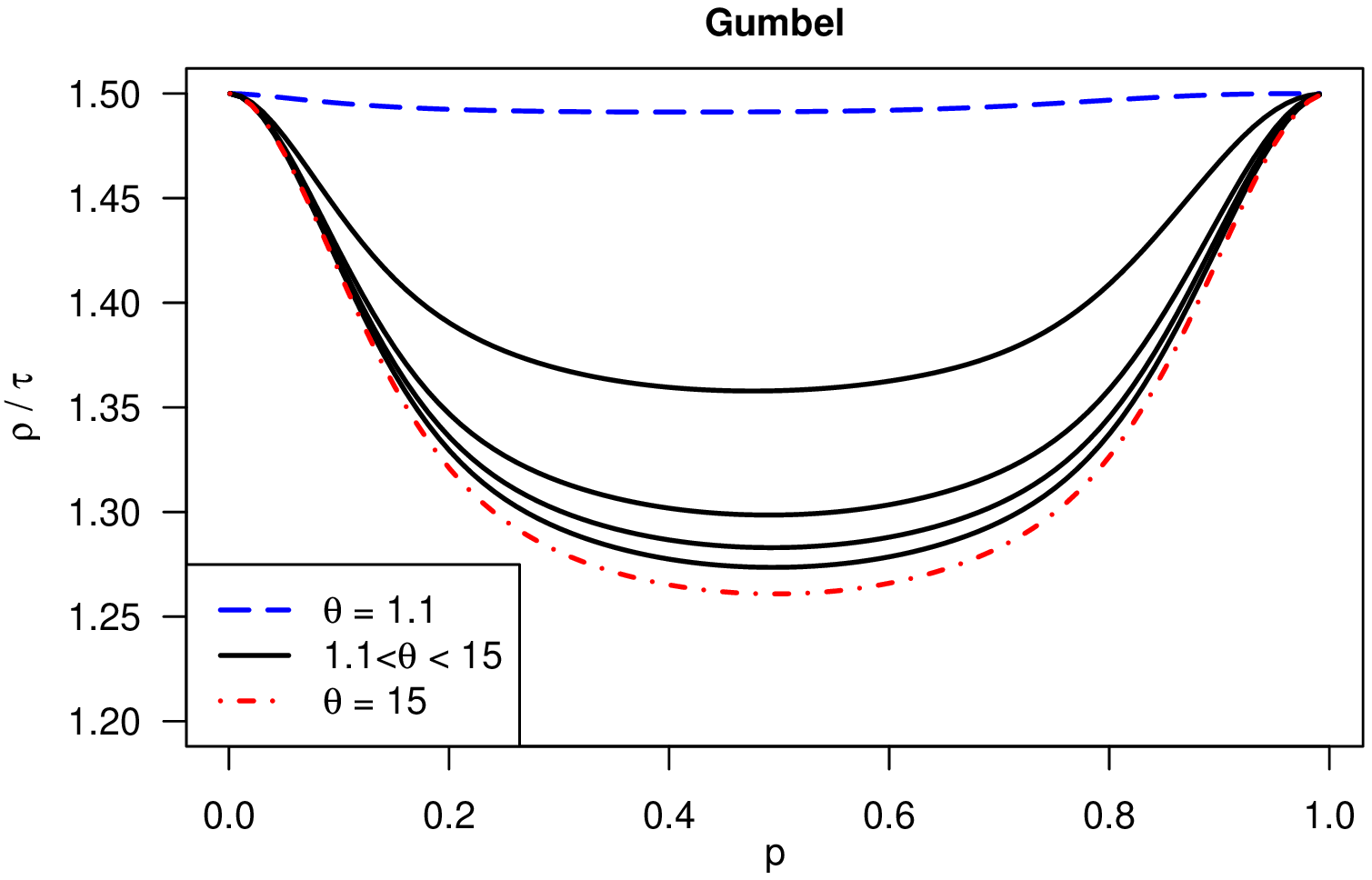}
        \end{subfigure}
    \vspace{-.4cm}
        \centering
 \begin{subfigure}[b]{0.34\textwidth}
                \includegraphics[width=\textwidth]{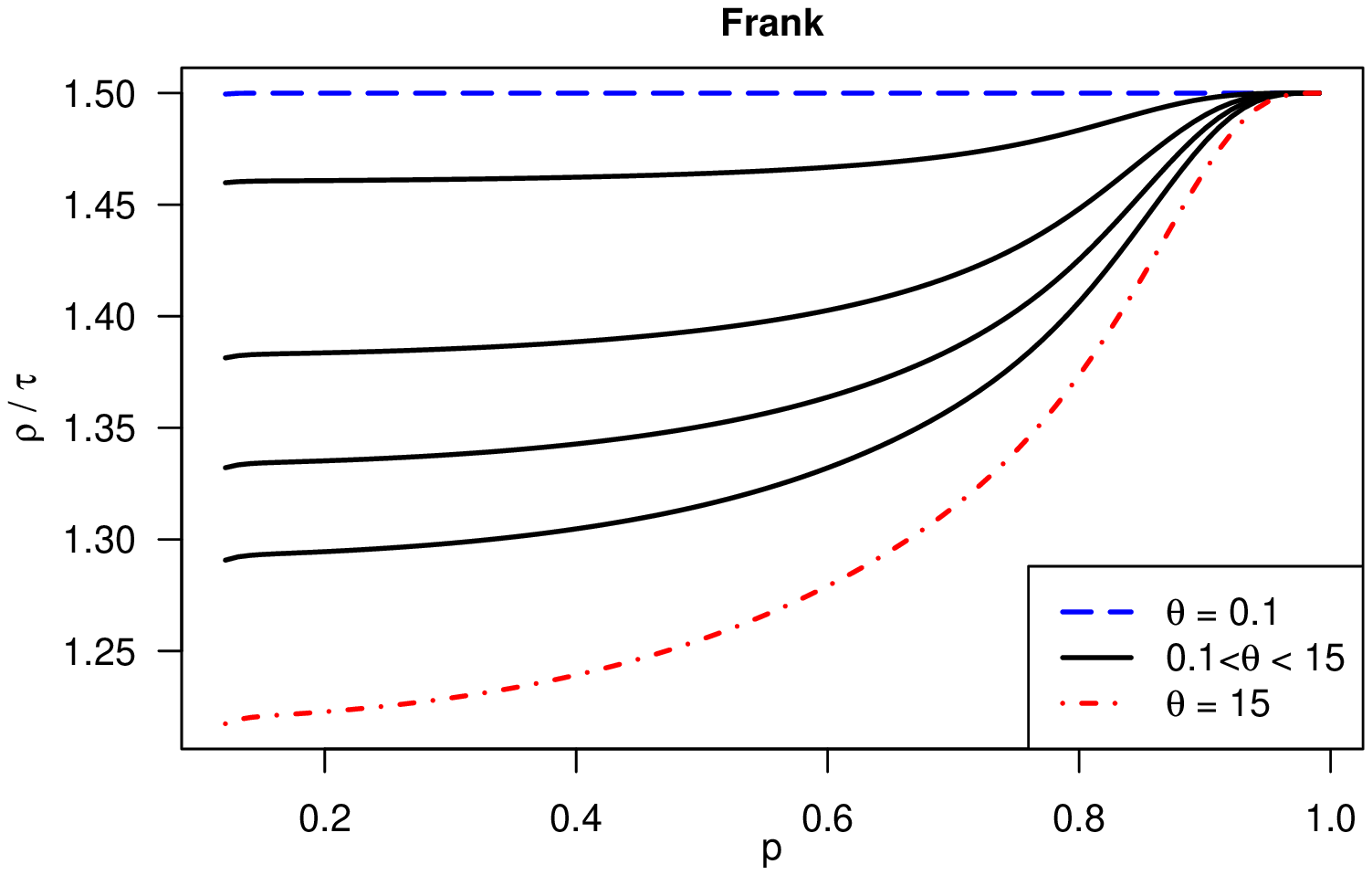}
        \end{subfigure}%
      \hspace{-.3cm}  
           \begin{subfigure}[b]{0.34\textwidth}
                \includegraphics[width=\textwidth]{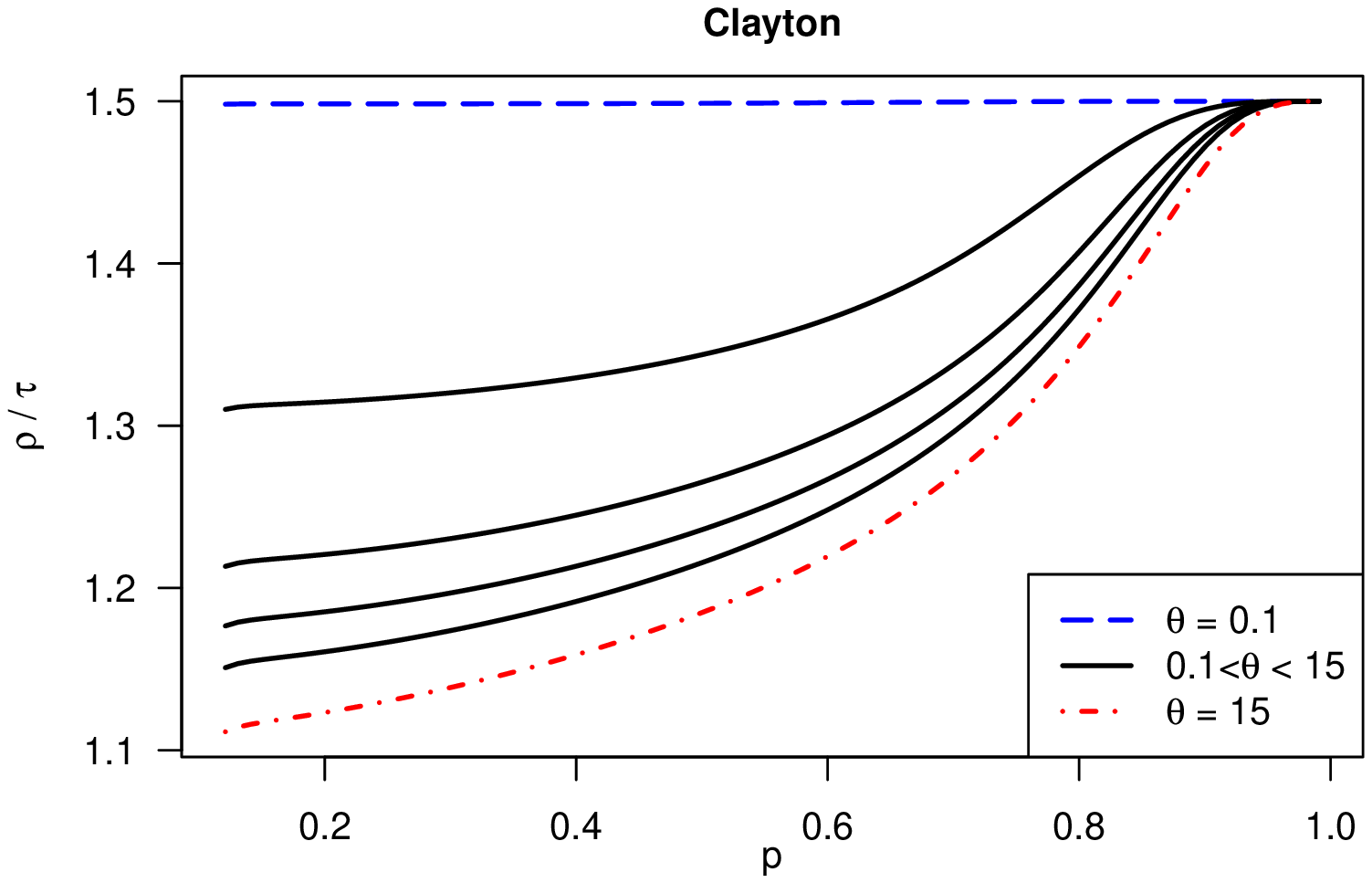}
        \end{subfigure}
         \hspace{-.4cm}
         \begin{subfigure}[b]{0.34\textwidth}
                \includegraphics[width=\textwidth]{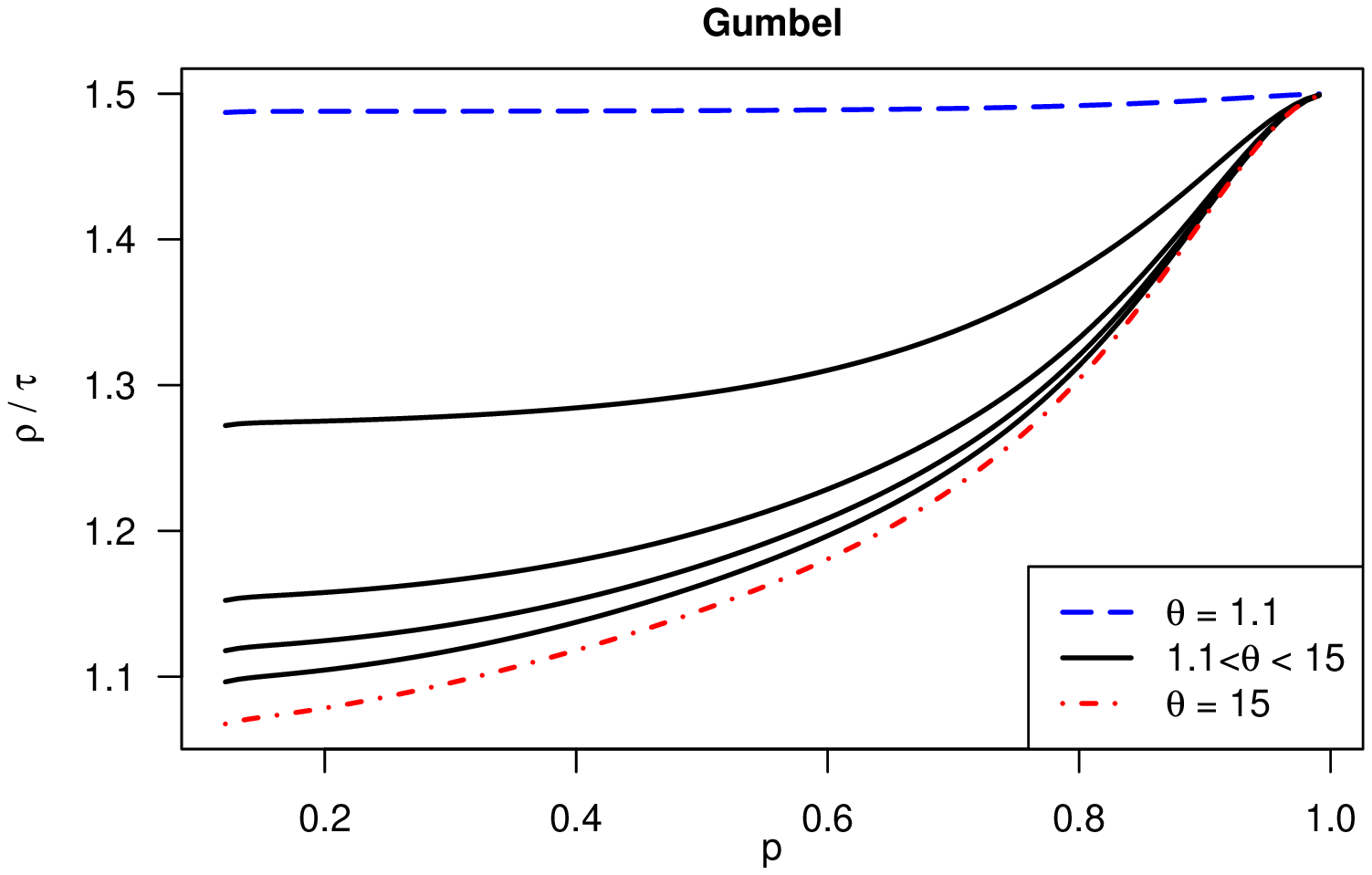}
        \end{subfigure}
    \vspace{-.1cm}
        \centering
 \begin{subfigure}[b]{0.34\textwidth}
                \includegraphics[width=\textwidth]{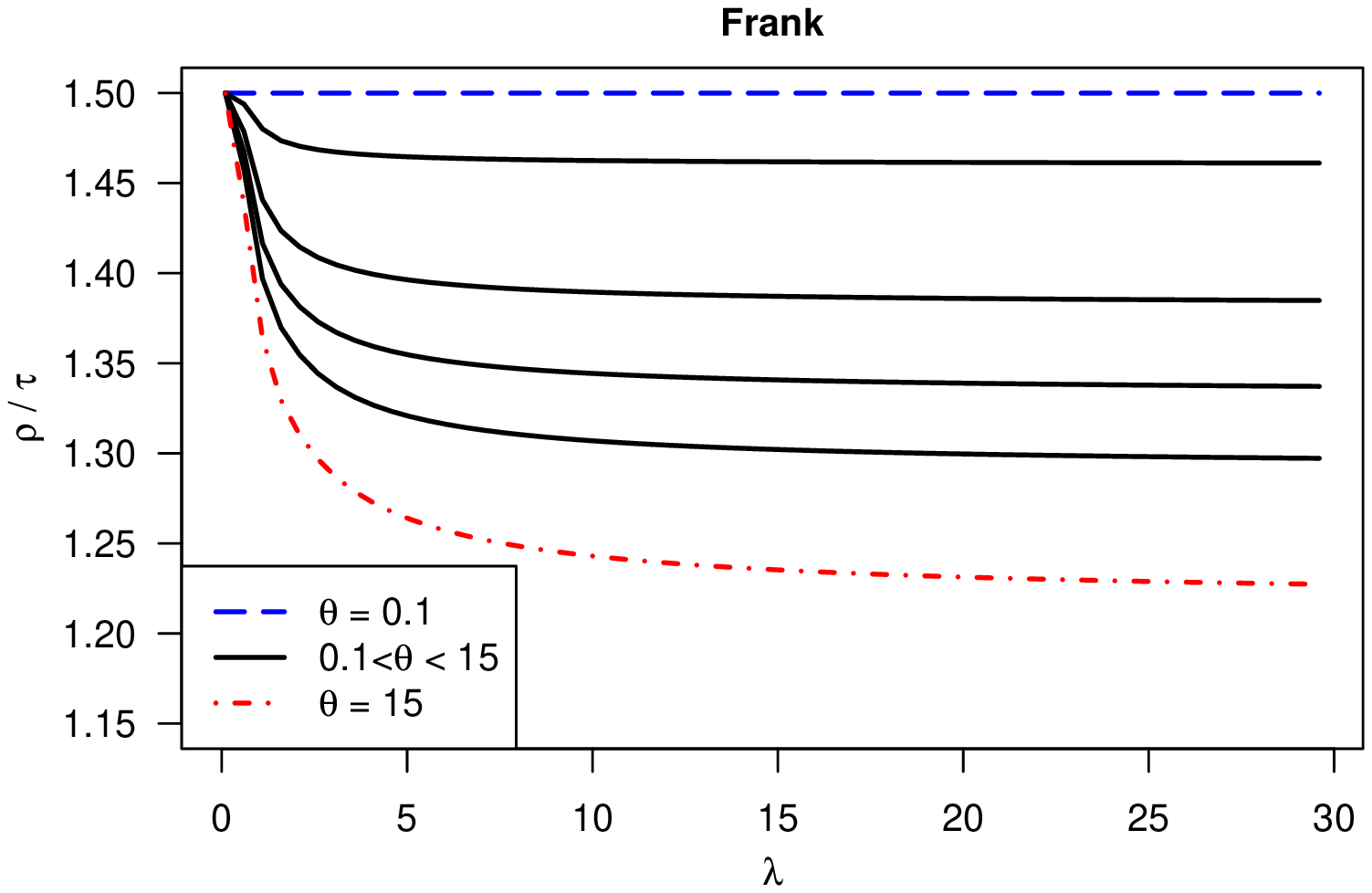}
        \end{subfigure}%
         \hspace{-.4cm}
           \begin{subfigure}[b]{0.34\textwidth}
                \includegraphics[width=\textwidth]{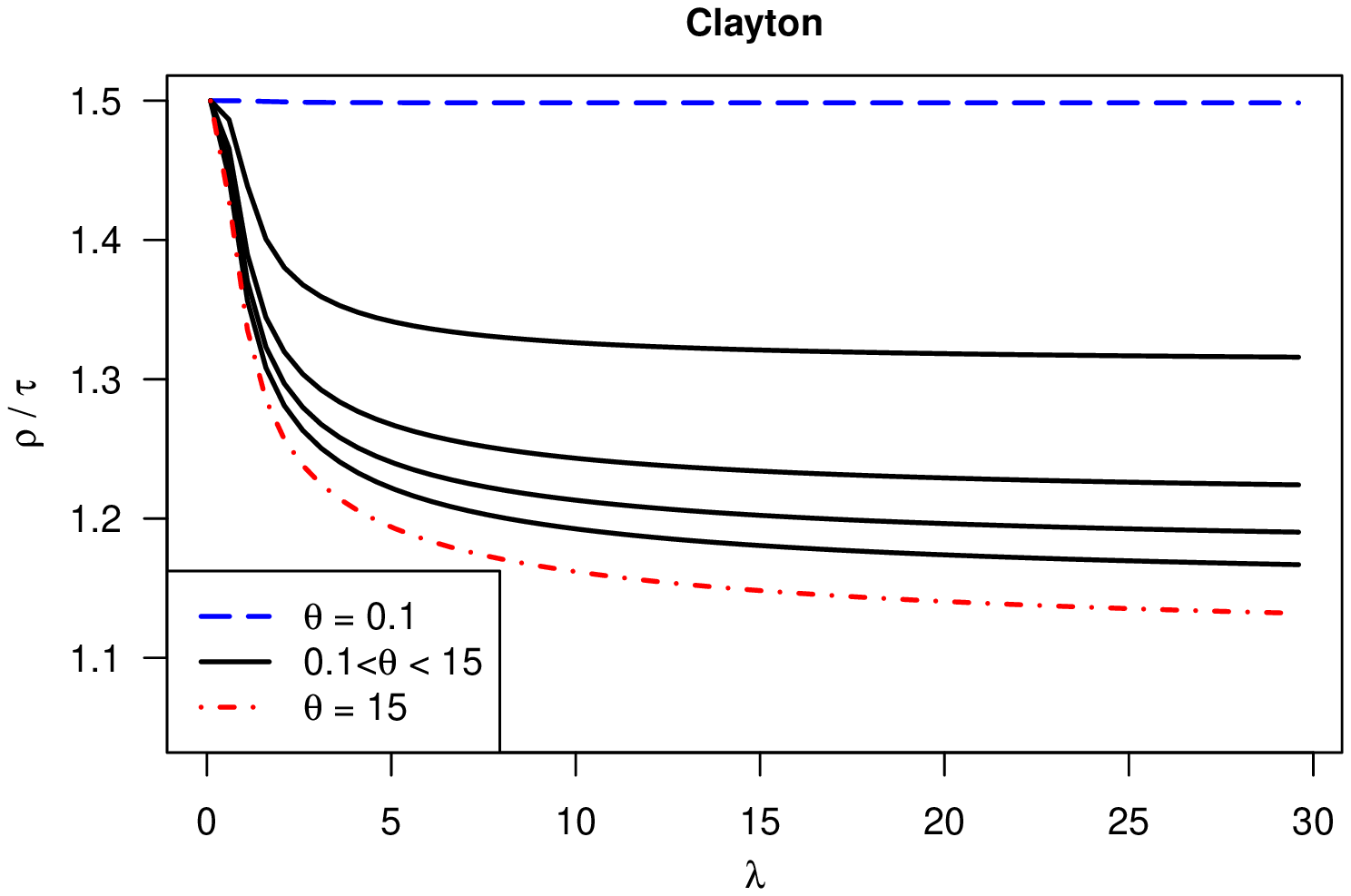}
        \end{subfigure}
          \hspace{-.4cm}
         \begin{subfigure}[b]{0.34\textwidth}
                \includegraphics[width=\textwidth]{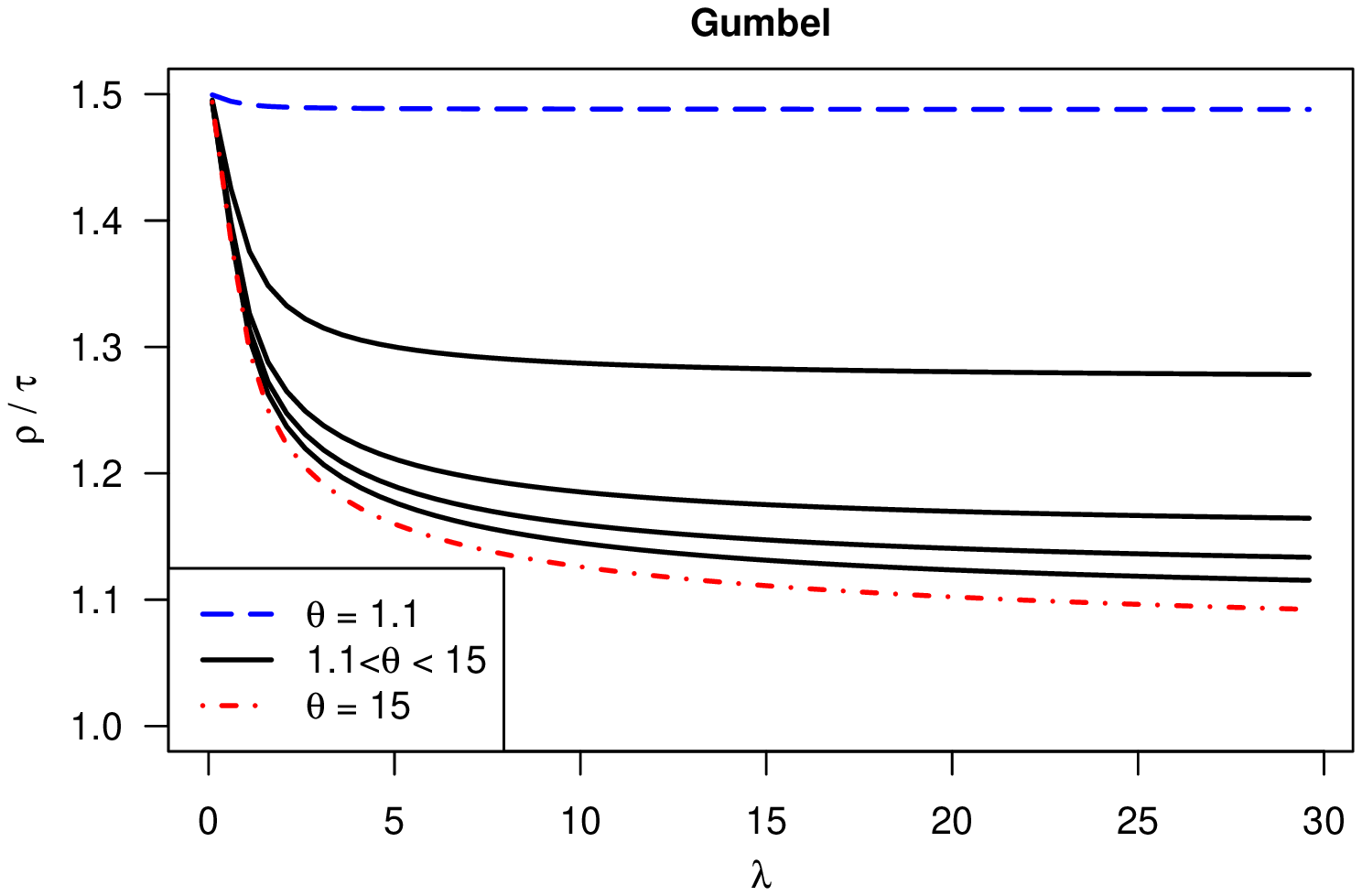}
        \end{subfigure}
  \caption{The ratio of  Spearman's rho to  Kendall's tau versus  the parameter of the marginals. The top row is for $Bin(5,p)$, the middle row is for $NB(4,p)$, and the bottom row is for $Poisson(\lambda)$.  }
  \label{f3}
  \end{figure}

\section{Real Data Analysis}

In this section, we illustrate the application of the  proposed copula models in practice by analyzing and measuring  the dependencies between different elements of \emph{Cervical Cancer} data set that  is gathered from a major hospital in Venezuela in the year 2017. 

\subsection{Data Characteristics}\label{subsec5.0}
  The  \emph{Cervical Cancer} data has been collected  from  ``Hospital Universitario de Caracas'' in Caracas, Venezuela in the year 2017 with a total of 667 patients.  The complete data set can be found at  \url{https://archive.ics.uci.edu/ml/datasets/Cervical+cancer+\%28Risk+Factors\%29}    

 \emph{Sexually transmitted diseases} (STDs) are venereal diseases which occur when pathogens are passed from one person to another by sexual activity.  Symptoms of STDs and infections usually appear and affect  the  genitalia and urinary tracts (Di Paolo, 2018). We refer to Loeper et al. (2018) for more details  about sexually transmitted diseases. We are interested in studying the relationship between the use of an intrauterine device (IUD) and the risk of STDs. The IUDs have been implicated in many studies in STDs. Summary of the frequency and percentages of patients based on their number of  years using IUD  and number of STDs diagnosed is presented in Table \ref{table4}.

\begin{table}[!htbp]
\caption{Frequency and percentages of the number of STDs diagnosed and the number of years of IUD use}\label{table4}
\begin{center}
\begin{tabular}{c | c c c c c}

 \backslashbox{ IUD$(Y)$}{STDs $(X)$}&    0 &  1  &  2  &  Total Number &    Percent  \\
 \hline
   0 &   537 &  25 &  30 &  592  &   88.75 \\
   1  &    36 &  0 &  6 &   42 &   6.30  \\
   2  &   22 &  2 &  1 &  25 &   3.75  \\
   3  &    4 &  0 &  1 &  5 &   0.75  \\
   4  &   3  &  0  &  0 &  3 &   0.45  \\
\hline
   Total Number  &   602  &  27 &  38  &  667 &   \\
   Percent  &   90.25  &  4.05 &  5.7  &   &   100  \\
\end{tabular}
\end{center}
\end{table}

Let $X_{i}$ and $Y_{i}$ represent the number of STDs diagnosed, and the number of years of IUD use  for patient $i$, respectively, for $i=1,2,\dots, 667$. Here, $X_{i}$ takes values $0, 1 ~\mbox{and}~ 2$, corresponding to the three groups of  number of STDs diagnosed.   Also, $Y_{i}$  takes values   $0, 1, 2, 3 ~\mbox{and}~ 4$, corresponding to the five groups of IUD users, ``not using IUD'', ``using IUD for less than 5 years'', ``using IUD between 5 and 10 years'', ``using IUD between 10 and 15 years'', and ``using IUD more than 15 years'', respectively.
   The results of  Table \ref{table4} show that
about 89\% (592 patients), prefer to not use IUD at all,
about 6\% (42 patients) use IUD for less than 5 years,
about 4\% (25 patients) use IUD between 5 and 10 years,
about 0.8\% (5 patients) use IUD between 10 and 15 years, and
about 0.5\% (3 patients) use IUD for more than 15 years.
These results are  not surprising.   The most common reasons that patients are not using IUD are  ``planned pregnancy'', ``lack of literacy'', ``lack of access to healthcare'', ``negative view of society'', or ``personal reasons'' (Petta et al., 1994).

In most of the patients (about 90\%), STDs are not diagnosed while about 10\%  of them are suffering from STDs.  Note that, there were 6 patients with more than 2 STDs who merged with the group of patients with 2 STDs and there was no patients with more than 4 STDs. Moreover,
among the 89\% of patients who did not use IUD, about 9.29\% had at least one STDs,
 among the 6.3\% patients who used IUD for less than 5 years, about 14.29\%   had at least one STDs.

\subsection{Specification of the Copula Model}\label{subsec5.1}
 We adopt a similar approach as in Zimmer and Trivedi (2006) and Shi and Valdez (2011) to estimate the  dependency structure of the cancer data.  Zimmer and Trivedi (2006) applied a trivariate copula to the model and jointly estimates the interdependence between  insurance decisions and health care demands among married couples, and Shi and Valdez (2011) used a bivariate copula to model the frequency of accidents and coverage selection in the automobile insurance market. 
  From a biostatistical perspective, Zhong and Cook (2016) used copulas to detect within-family associations in   chronic diseases data. 
   In this study, we apply a bivariate copula to model and estimate the joint distribution to find the effect of the number of years of IUD use  on the number of STDs.

Parametric copula functions are used to estimate the joint probability mass function $X$ and $Y$. The first step in the copula approach is to specify the marginal distributions. In this study, the marginal variables  $X$ (the number of STDs) and $Y$ (the number of years of IUD use)  are non-negative integer count variables. We considered both \emph{Poisson} and \emph{Negative Binomial} distributions to fit the marginal variables $X$ and $Y$. The goodness-of-fit test rejected the \emph{Poisson}  assumption  for the marginal data. However, the goodness-of-fit test indicated that the \emph{Negative Binomial-2}  distribution, $NB_2(\mu, \psi)$, where $\mu$ is the mean and $\psi$ denotes the overdispersion parameter, fits the marginal data well.  The probability mass function of  $NB_2(\mu, \psi)$ is given in Eq \eqref{eq5.1}.    See the results of the goodness-of-fit tests in Section \ref{subsec5.2}.   Therefore, we specify $F_1(t_1)$ and $F_2(t_2)$ as CDFs of \emph{Negative Binomial-2} distribution,  where $F_1(\cdot)=F_X(\cdot)$ and $F_2(\cdot)=F_Y(\cdot)$.  This specification  provides a flexible framework for count data regression analysis. For each observation $i = 1,2, \dots, 667$, each marginal is defined conditionally on  a set of covariates ${\bf Z}_i$ with corresponding parameter vectors $\boldsymbol\beta_1 $ and  $\boldsymbol\beta_2$. That is,
 \begin{equation}\label{eq5.1}
  F_j(t_{ij}| {\bf Z_i},\boldsymbol\beta_j )= \sum_{k=0}^{t_{ij}}{\psi_j +k -1\choose k} \left( \frac{\psi_j }{\mu_{ij} + \psi_j} \right)^{\psi_j} \left( \frac{ \mu_{ij} }{\mu_{ij} + \psi_j}\right)^k, ~~ j=1,2,~~i=1,2,\dots, 667,
 \end{equation}
 where
 \begin{equation}\label{eq5.2}
 E(X_{i}| {\bf Z}_{i}) =\mu_{i1} = \exp({\bf Z}^{'}_{i} \boldsymbol\beta_1), ~~~~~~~ E(Y_{i}| {\bf Z}_{i}) =\mu_{i2} = \exp({\bf Z}^{'}_{i} \boldsymbol\beta_2),
 \end{equation}
 are the conditional means, and  their conditional variances  are given by  $\mu_{ij}\left(1+ \mu_{ij}/\psi_j \right)$, for $j=1,2$.
  That is, the covariates are incorporated into the  model via a log link function.
 Here, the covariates refer to certain variables or information related to the patients such as  age, smoke status, etc. All of the covariates are listed in Table \ref{Table6}.

  After specifying the marginal distributions, the unknown joint distribution function of $X$ and $Y$ can be constructed  by using  an appropriate copula function as follows
  \begin{equation}\label{eq5.3}
   {H}({\bf t};  \boldsymbol\beta_1 ,\boldsymbol\beta_2, \theta)  = \mathcal{C} \left( F_1(t_{i1}| {\bf Z_i},\boldsymbol\beta_1 ), F_2(t_{i2}| {\bf Z_i},\boldsymbol\beta_2 ); \theta  \right).
  \end{equation}

  The method of inference function for margins (IFM) is applied to estimate the parameters of the proposed model in Eq \eqref{eq5.3}. The IFM  approach is a  two-step procedure that  proposed by Joe (1997), and  McLeish and Small (1988).  At the first step, the parameters of the marginal distributions are estimated by maximizing the following marginal log-likelihood  functions
  \begin{equation}\label{eq5.4}
   L_{X}(\boldsymbol\beta_1) =\sum_{i=1}^{n} \log f_{X}(x_{i},\boldsymbol\beta_1),~~~~ L_{Y}(\boldsymbol\beta_2) =\sum_{i=1}^{n} \log f_{Y}(y_{i},\boldsymbol\beta_2),
  \end{equation}
  where $f_X(\cdot)$ and $f_Y(\cdot)$ are the pmf of $X$ $Y$, respectively.   At the second step, each parametric margin is substituted into the following copula likelihood function as
  \begin{equation}\label{eq5.5}
     L(\theta;\widehat{\boldsymbol\beta}_1,\widehat{\boldsymbol\beta}_2)=\sum_{i=1}^{n} \log h(x_{i},y_{i},\widehat{\boldsymbol\beta}_1,\widehat{\boldsymbol\beta}_2;\theta),
  \end{equation}
 where $h(\cdot, \cdot)$ is the joint pmf of $X$ and $Y$ defined in Eq \eqref{eq3.2}.  Then, this joint log-likelihood is maximized with respect to the copula parameter $\theta$. Note that, the IFM method computationally is more feasible than the full maximum likelihood approach. Moreover, the IFM estimators are consistent and asymptotically normal (Joe, 2005).

 \subsection{Estimation Results and Discussion}\label{subsec5.2}

 Goodness-of-fit tests are carried out for the marginal variables STDs and IUD. Both the \emph{Poisson} and \emph{Negative Binomial} distributions are fitted to the  marginal data.  If we fit a Poisson($\lambda$) distribution to STDs, then the MLE of $\lambda$ is $\bar{X}_n= 0.1544$, the chi-square goodness-of-fit test statistic is $17.928$ with the p-value $0.0001$. Similarly,  for IUD, the MLE of $\lambda$ is $\bar{Y}_n= 0.1783$, the chi-square goodness-of-fit test statistic is $11.489 $, and  the p-value is  $0.0216$.  Therefore, the null hypotheses  that the STDs or IUD come from a Poisson distribution are rejected.  However, if we fit a \emph{Negative Binomial-2} distribution, $NB_2(\mu, \psi)$, the results of goodness-of-fit tests show that it fits  both the STDs and IUD well. The results of chi-square goodness-of-fit tests for the \emph{Negative Binomial-2}  distribution with the   observed and fitted frequencies of  the STDs and IUD  are presented  in Table \ref{table12}. Moreover, the null hypothesis  that the data fit a zero inflated model is rejected for both  variables STDs and IUD.

 \begin{table}[!htbp]
\caption{ Goodness-of-fit tests of the \emph{Negative Binomial-2}  model for both margins }\label{table12}
\begin{center}
{\small
\begin{tabular}{ l l l l l l l l l l l l l}
  \hline
 &  &  &\multicolumn{3}{  l}{STDs}& & && &\multicolumn{3}{l}{IUD}\\
  \cline{1-6}   \cline{8-13}
&\multicolumn{2}{c}{Observed}&&\multicolumn{2}{c}{Fitted}&&&\multicolumn{2}{c}{Observed}&&\multicolumn{2}{c}{Fitted}\\
\cline{2-3}  \cline{5-6}   \cline{9-10}  \cline{12-13}
Value &  \%     &  Count      &&  \%    &   Count    && Value   &  \%         & Count   && \%         &  Count  \\
\cline{1-6}\cline{8-13}
0       &  90.25   &   602      &&  90.07  &   600.77  &&   0       &    88.76     &   592    &&  88.64    &   591.23 \\
1       &  4.05    &    27        && 6.67    &  44.49     &&   1       &    6.30        &   42      &&  7.55     &   50.36   \\
2       &  5.70      &   38         && 3.26    & 21.74    &&   2       &    3.75      &   25      &&   2.34     &   15.61   \\
\cline{1-6}
 \multicolumn{6}{ c }{ $\hat{\mu}=\bar{X}_n= 0.1544$ ~~~  $\hat{\psi}= 0.1421$   }        &&   3       &    0.75      &    5      &&    0.99     &   6.60    \\
   \multicolumn{6}{ c }{    chi-square=3.6882 ~~~~   p-value=0.1582   }        &&   4       &    0.45      &    3      &&    0.48 &   3.2   \\
   \cline{8-13}
    &      &        &&          \multicolumn{9}{ c }{ ~~~~~~~~~~~~~~~~~~$\hat{\mu}= \bar{Y}_n= 0.1785$ ~~~  $\hat{\psi}= 0.1630 $}\\
    &      &        &&      &        &&          \multicolumn{6}{ c }{    chi-square=0.7546 ~~~~   p-value=0.9444 }     \\
\hline
\end{tabular}}
\end{center}
\end{table}

\begin{sidewaystable}[!htbp]
\caption{Descriptive statistics of the covariates used in the model calibration}\label{Table6}
\scriptsize
\begin{center}
\begin{tabular}{l l l l l l l l l l l l l l l l l}
\cline{1-12}
\multicolumn{17}{l}{(a) Covariates used for modeling the number of STDs}\\
\cline{1-12}
&&&&\multicolumn{2}{c}{$STDs$: 0} && \multicolumn{2}{c}{$STDs$: 1} && \multicolumn{2}{c}{$STDs$: 2} &&& \\
  \cline{5-6}   \cline{8-9}    \cline{11-12}
  Variable   &    M(\%)  &   Std &&  M(\%) &    Std  &&  M(\%) &    Std &&  M(\%) &    Std &&  &   &&         \\
  \cline{1-12}
Smoke=1, if patient smokes, 0 if not &   14.24 & 0.35  & &  12.79 &   0.334  &&   29.63  &   0.465 &&   26.32  &   0.446 & && &&\\
 Age=1, if patient's age is less than 25 $^1$ & 43.93 & 0.497 && 44.19 & 0.497 && 29.63  &  0.465  &&  50  &  0.507 & &&&& \\
Age=2, if patient's age is between 25 and 45 &   52.92 &  0.499 & & 52.49  & 0.5 && 66.67 &  0.48  && 50 & 0.507 &&&&&\\
Age=3, if patient's is 45 or more &  3.15  &  0.175 & & 3.32  &  0.18  &&  3.7   &  0.192  &&  0  &   0
&   & && &\\
HC=0, if patient didn't used hormonal contraceptives $^1$ &  35.53  &  0.479 & & 35.05  &  0.478   && 40.74    &   0.501 &&  39.47  &  0.495   && && &\\
HC=1, if patient used hormonal contraceptives for less than 10 years &   59.07 &  0.492 & &  59.63 &   0.491  &&  51.85   &  0.509  &&  55.26  & 0.504    &  & && &\\
HC=2, if patient used hormonal contraceptives for 10 years or more &   5.4 &  0.226  &&  5.3 & 0.225    &&   7.41  & 0.267   && 5.26   &  0.226   && && &\\
AFS=1, if age of patient is less than 15 at the time of first sexual intercourse $^1$ &  11.40  & 0.318  && 11.30  & 0.317    &&  14.81   & 0.362   && 10.53   &  0.311   && && &\\
AFS=2, if age of patient is 15, 16 or 17 years at the time of first sexual intercourse  & 50.97   &  0.5 &&  51 & 0.5    && 59.26    & 0.501   && 44.74   & 0.504    && && &\\
AFS=3, if age of patient is 18 years or more at the time of first sexual intercourse  & 37.63   &  0.485 && 37.71  &  0.485   && 25.93    & 0.447   &&  44.74  &  0.504   && && &\\
NSP=1, if the number of sexual partners are 1 or 2 $^1$ &  56.52  & 0.496  & &  57.31 &  0.495   &&  29.63   &  0.465  && 63.16   &  0.489   & & && &\\
NSP=2, if the number of sexual partners are 3 or 4 &   35.38 & 0.479  && 34.88  & 0.477    &&   59.26  &  0.501  && 26.32   &   0.446  && & & &\\
NSP=3, if the number of sexual partners are 5 or 6 &  6.75  & 0.251  & &  6.64 &   0.249  &&  11.11   &  0.32  && 5.26   &  0.226   &&& && \\
NSP=4, if the number of sexual partners are 7 or more & 1.35   & 0.115  & & 1.16  &  0.107   &&  0   &   0 &&  5.26  &  0.226
&&   & && \\
NP=0, if patient didn't had any pregnancy $^1$ & 2.1 &  0.143 && 2.16 & 0.145 &&  3.7 & 0.192  && 0 & 0
  &&& && \\
NP=1, if the number of pregnancies are 1,2,3 or 4 &   89.81 &  0.303  && 89.87  &   0.302  &&   77.78  &  0.424  &&  97.37  &  0.162   && && &\\
NP=2 if $\#$ of pregnancies are 5 or more &  8.1  &  0.273 & & 7.97  &   0.271  &&   18.52  &  0.396  &&   2.63 &  0.162   &&& &&
\end{tabular}
\end{center}
\begin{center}
\begin{tabular}{l l l l l l l l l l l l l l l l l}
\cline{1-15}
\multicolumn{15}{l} {(b) Covariates used  for modeling the number of years of IUD use } \\
\cline{1-15}
& \multicolumn{2}{c} { $IUDY$: 0} && \multicolumn{2}{c} {$IUDY$: 1}  & & \multicolumn{2}{c} {$IUDY$: 2} &&\multicolumn{2}{c}    {$IUDY$: 3}  & & \multicolumn{2}{c}   {$IUDY$: 4}\\
 \cline{2-3}   \cline{5-6}    \cline{8-9}   \cline{11-13} \cline{14-15}
   Variable   &  M(\%) &    Std  &&  M(\%) &    Std &&  M(\%) &    Std &&  M(\%) &    Std &&  M(\%) &    Std \\
  \cline{1-15}
Smoke=1     &  14.86 &   0.356  &&   9.52  &  0.297   &&  8   &   0.277  &&  20  &  0.447 && 0  & 0 \\
Age=1 $^1$     &  48.14 &   0.5  &&   14.29  &   0.354  &&   8  &   0.277  &&   0 &  0  && 0  &  0  \\
  Age=2     & 49.16 & 0.5 && 80.95 & 0.397  &&   84  &   0.374&&   80 &  0.447 &&   100 & 0  \\
  Age=3     &  2.7 &   0.162  &&   4.76 &   0.216  &&   8 &   0.277  & &  20 &  0.447 && 0 & 0 \\
   HC=0 $^1$  & 36.32 & 0.4813  &&  16.67 & 0.377 && 40 & 0.5 & & 60 & 0.548  && 66.67  & 0.577 \\
    HC=1   & 59.29 & 0.492 && 64.29 & 0.485 && 52 & 0.51  & & 40 &  0.548 && 33.33  & 0.578 \\
   HC=2  & 4.39 & 0.205   &&  19.05 &  0.397  &&  8  & 0.277  & & 0  &   0  && 0  & 0 \\
AFS=1 $^1$  &11.15  & 0.315   && 11.9  &  0.328  && 12   & 0.332 &&  20 & 0.447    &&  33.33 & 0.577 \\
 AFS=2  & 51.01 & 0.5   && 50  &   0.506 &&  52  & 0.51  && 40  & 0.548    && 66.67  & 0.577\\
 AFS=2  & 37.84 & 0.485   && 38.1  & 0.492   && 36   & 0.49  && 40  & 0548    && 0 & 0\\
 NSP=1 $^1$   & 58.11 & 0.494 && 40.48 & 0.497 && 48 & 0.51 && 40 & 0.548  && 66.67 & 0.578 \\
 NSP=2  & 33.61 &  0.473  &&  50 &  0.506  &&  48  &  0.51 && 60 & 0.548 && 33.34 & 0.578 \\
 NSP=3   & 6.93 &  0.254  && 7.14  &  0.261  && 4   &  0.2 &&  0 &  0   &&0   & 0 \\
  NSP=4   & 1.35 &  0.116  &&  2.38 &  0.154  &&  0  &  0 && 0  &   0  &&  0 & 0 \\
  NP=0 $^1$  & 2.36 &  0.152  && 0  &  0  &&  0  & 0  && 0 &  0   && 0 & 0 \\
  NP=1 & 90.71 &  0.291  &&  83.33 & 0.377   &&  84  &  0.374 &&  60 &  0.548   &&  100 & 0 \\
  NP=2 & 6.93 &  0.254  &&  16.67 &  0.377  &&  16  & 0.374  &&  40 &   0.548  && 0  & 0 \\
  \cline{1-15}
\footnotesize{$^1$ reference level}\\
\end{tabular}
\end{center}
\end{sidewaystable}

The covariates used in this  study  are  presented in Table \ref{Table6}.  The covariates included  demographic characteristics and medical conditions  such as  age, smoke status,  using or not using hormonal contraceptives (HC), age at first sexual intercourse (AFS), number of sexual partners (NSP), and  number of pregnancies (NP). Note that, the same covariates are used for both margins, i.e., the number of STDs and the number of years of IUD use. Moreover, all of the covariates (explanatory variables) are categorical variables.  Descriptive statistics of the  covaritates are presented in Table \ref{Table6} (a) and (b).

The generalized negative binomial regression model defined in Eq  \eqref{eq5.2} is fitted to the data.
 Table  \ref{table7} shows  the  estimation results of the parameters, $\widehat{\boldsymbol\beta}_1$, corresponding to the regression model defined in Eq  \eqref{eq5.2} for margin $X$ (STDs).    Similarly, Table \ref{table9} provides  the estimation results, $\widehat{\boldsymbol\beta}_2$, for   margin  $Y$ (IUD). 
The analysis shows that  patient's age at first sexual intercourse (AFS) is an important factor that is associated with IUD.   In a different study,   Ethier et al., (2018) has also  shown  that the AFS is an important and significant covariate on sexually transmitted diseases (STDs). Note that, in our study, the AFS is categorized as $<15$, $15-17$, and $\geq 18$ years.

  Although there is no information about the marital status of the patients in our study, some studies have indicated that married individuals are possibly more open-eyed and attentive about their sexual activities. For instance, Finer et al. (1999) demonstrated that the risk of STDs for unmarried women is more than for cohabiting women, and the cohabiting women  are more likely than currently married women to be at risk.

\begin{table}[!htbp]
\caption{Estimates of the NB model for STDs with all covariates}\label{table7}
\begin{center}
{\small
\begin{tabular}{l l l l}
\hline
STDs-NB       &   Estimate({ $\hat{\boldsymbol\beta}_1$})   &  StdDev     & $p$-value \\
\hline
Intercept & -2.5964  &  1.2307  &   0.0349  \\
Smoke    &   0.8070  &  0.3729   &   0.0304  \\
Age=2    &    -0.0651  &  0.3264  &   0.8419   \\
Age=3    &    -1.1424  &  1.1699  &   0.3288   \\
HC=1     &   -0.1954  &  0.3002  &   0.5153   \\
HC=2    &     0.1351  &  0.6518   &   0.8357   \\
AFS=2     &    0.0731  &  0.4729   &   0.8772   \\
AFS=3     &    0.2263  &  0.5112  &   0.6580   \\
NSP=2    &    -0.0018  &  0.3184  &   0.9956   \\
NSP=3    &     0.0657  &  0.5723   &   0.9086   \\
NSP=4    &     0.9982  &  1.0007  &   0.3185   \\
NP=1      &    0.5948  &  1.1894   &   0.6170   \\
NP=2     &     0.4197  &  1.3055   &   0.7479   \\
Dispersion &  0.1660 &&\\
\hline
\multicolumn{2}{ l }{AIC= 589.43}      &  \multicolumn{2}{ l }{-2log-Like.=561.434}  \\
\hline
\end{tabular}}
\end{center}
\end{table}
\begin{table}[!htbp]
\caption{Estimates of the NB model for STDs, after excluding the non-significant covariates}\label{table8}
\begin{center}
{\small
\begin{tabular}{l l l l}
\hline
STDs-NB           &   Estimate({ $\hat{\boldsymbol\beta}_1$})   &  StdDev &   $p$-value  \\
\hline
Intercept   & -2.0317  &   0.1567   &   0.0000 \\
Smoke      &  0.8100  &   0.3576   &  0.0235  \\
Dispersion & 0.1557  &   &\\
\hline
\multicolumn{2}{l}{AIC =   570.87}      & \multicolumn{2}{l}{-2log-Lik. = 564.865}      \\
\hline
\end{tabular}}
\end{center}
\end{table}

   Simple linear regression and stepwise regression analysis are used to identify the significant covariates in the generalized negative binomial regression model defined in Eq \eqref{eq5.2} for both STDs and IUD responses. The results of the stepwise regression analysis for the STDs and IUD  are summarized in Table \ref{table8} and Table \ref{table10}, respectively. As the result, \emph{Smoke status} is the only significant covariate in the model of STDs whereas \emph{Age} and \emph{AFS}  are the significant covariates in the model of IUD. Moreover, intercept is significant in both cases.

\begin{table}[!htbp]
\caption{Estimates of the NB model for IUD with all covariates}\label{table9}
\begin{center}
{\small
\begin{tabular}{l l l l}
\hline
IUD-NB          &   Estimate({ $\hat{\boldsymbol\beta}_2$})        &  StdDev         &  $p$-value \\
\hline
Intercept   & -29.1300  & 193400   & 0.9999    \\
Smoke      &  -0.7540   & 0.4080  & 0.0646 \\
Age=2      &   2.4580  & 0.4043   & 0.0000 \\
Age=3      &   2.6110   & 0.6959   & 0.0001 \\
HC=1      &  -0.3450 & 0.2736    & 0.2073  \\
HC=2      &  -0.2972 & 0.4763  & 0.5326    \\
AFS=2      &  -0.7378 & 0.4221  & 0.0804 \\
AFS=3      &  -1.3060 & 0.4552  & 0.0041 \\
NSP=2     &    0.1421 &  0.2681   & 0.5959   \\
NSP=3      &  -0.8216 & 0.5913  & 0.1647    \\
NSP=4      &  -0.5206 & 1.145  & 0.6493    \\
NP=1       &   26.64 & 1.934   & 0.9999    \\
NP=2        &  26.90 & 193400   & 0.9999 \\
Dispersion       &   0.3630           &                     &                 \\
\hline
   \multicolumn{2}{l}{AIC = 592.11}              &    \multicolumn{2}{ l }{-2log-Lik.= 564.11}          \\
\hline
\end{tabular}}
\end{center}
\end{table}

 After estimating the parameters of the marginal distributions by maximizing the likelihood functions defined in Eq \eqref{eq5.4}, the second step of the IFM method, described in Section \ref{subsec5.1}, is applied to estimate the parameters of the joint model. To this end, different copula functions are used to estimate the population version of Kendall's tau and Spearman's rho between STDs and IUD marginal variables. The estimation results are presented in Table \ref{table13}.
 We first  consider the Frank copula  due its  versatility and flexibility to model  both positive and negative dependencies. The dependence parameter of the Frank copula $\theta$, is estimated  to be $0.93854$ which resulted  in a Spearman's rho of  $0.0095$. Similarly, all of the  results in Table \ref{table13} indicate a very weak positive relationship between usage of IUD and the number of STDs.

 \begin{table}[!htbp]
\caption{Estimates of the NB model for IUD, after excluding non-significant covariates}\label{table10}
{\small
\begin{center}
\begin{tabular}{l l l l}
\hline
IUD-NB              &   Estimate({ $\hat{\boldsymbol\beta}_2$})  &  StdDev  &   $p$-value  \\
\hline
Intercept  & -2.7306   &   0.4330   & 0.0000\\
AGE=2        &   2.4043    &  0.3859    &   0.0000 \\
AGE=3        &   2.7176   &   0.6282    &  0.0000  \\
AFS=2         & -0.7880   &   0.4222   &  0.0620 \\
AFS=3         & -1.2700   &   0.4450   &  0.0043 \\
Dispersion          &    0.3055    &              &               \\
\hline
\multicolumn{2}{ l }{AIC =   588.18}  & \multicolumn{2}{ l }{-2log-Lik.=  576.18}       \\
\hline
\end{tabular}
\end{center}}
\end{table}


\begin{table}[!htbp]
\caption{Estimates of copula parameters,  Kendall's tau, and  Spearman's rho of IUD and STDs}\label{table13}
{\footnotesize
\begin{center}
\begin{tabular}{c c c c c}
\hline
Family    &  $\hat{\theta}$ & -2Log-Lik. &   $\hat{\tau}(X,Y)$     &   $\hat{\rho}^S(X,Y)$    \\
\hline
Frank     &   0.9338     & 1139.702   & 0.0063   &  0.0095   \\
Clayton   &     0.4318   &  1139.879  &  0.0056  &  0.0084  \\
Gumbel  &   1.0502       & 1138.152   &  0.0089  &  0.0133 \\
Ali-M-H   &     0.4653   &  1139.790  & 0.0058   &  0.0086   \\
Joe        &    1.0598     &  1138.021  &  0.0089  & 0.0134   \\
\hline
\end{tabular}
\end{center}}
\end{table}

 There are several discussions in the literature which  conclude that using poorly designed IUD made women more vulnerable to the infections and STDs in 1970s and after,  and as a result  some women who using it died due to severe infections. However, after 50 years or so, the design of IUDs is vastly improved, and  therefore we expect that although
 the IUD does not protect against STDs but the modern IUDs themselves do not induce or accelerate the STDs.

Another way to assess the effect of usage of IUD on the number of STDs is to compare the conditional expectations of the number of STDs ($X$) given the number of years an IUD used ($Y$). To this end, first we compute the conditional probability of the number of STDs given the  IUD status for each patient by
      \begin{equation}\label{eq5.6}
P\left(X_{i}=x_{i}|Y_{i}=y_{i} \right)= f_{X_i|Y_i}\left(x_{i}| y_{i}; {\bf z}_1,  {\bf z}_2 \right) = \dfrac{ f \left(x_{i}, y_{i}| {\bf z}_1,  {\bf z}_2 \right)}{ f \left( y_{i}|   {\bf z}_2 \right)},
 \end{equation}
 where ${\bf z}_1$ and ${\bf z}_2$ are the significant covariances of  STDs and IUD given in Table \ref{table8} and Table  \ref{table10}, respectively.
  Then, given each IUD status, these probabilities are aggregated.   The  results are summarized in Table \ref{table14}.

\begin{table}[!htbp]
{\small
\caption{Conditional probability of the number of STDs  given the  IUD status (StdDev) }\label{table14}
\begin{center}
\begin{tabular}{c c c c c c   }
\hline
STDs&   {$IUD=0$}&  {$IUD=1$}& {$IUD=2$}& {$IUD=3$}&  {$IUD=4$}\\
  \hline
 0 &    0.9067	 (0.0203)&  0.8745 (0.0351) &  	0.8418	 (0.0473)	&  0.8249    (0.0377)	&  0.8137  	(0.0382)\\
1  &   0.0689	 (0.0104)&  0.0853 (0.0119) &  	0.0944	 (0.0102)	&  0.1000    (0.0081)	&  0.1022  	(0.0061)\\
2   &    0.0244	 (0.0069)&  0.0402 (0.0110) &  0.0638	 (0.0106)	&  0.0751	  (0.0097)	&  0.0841	 (0.0112)\\								
\hline
\end{tabular}
\end{center}}
\end{table}

Then, we used the conditional probabilities provided in  Table \ref{table14} to compute the desired conditional expectations. Particularly, we compute and compare  the difference in the conditional expectations, i.e.,  $E(X|Y=j) - E(X|Y=j-1)$, $j=1,2,3,4$, to investigate whether increased  use of IUD makes women  more vulnerable to STDs. The results are summarized in Table \ref{table15}. As we expected from the results of Spearman's rho and Kendall's tau in Table \ref{table13}, all of the differences in expectations  are very small. That is,  the effect of IUDs on STDs statistically is not significant.

\begin{table}[!htbp]
\caption{The effect of the number of years of IUD use on the number of STDs}\label{table15}
{\small
\begin{center}
\begin{tabular}{c c c c c c }
\hline
	           &  Mean     &  StdDev & 1st Quartile  & 2nd Quartile		& 3rd Quartile	\\	
\hline
	$E(X|Y=1) - E(X|Y=0)$	 &  0.0565   &  0.0408  &  0.0420 &  0.0565 &  0.0709  	\\			
	$E(X|Y=2) - E(X|Y=1)$      &   0.0739 &  0.0417 &  0.0592 &  0.0739 &  0.0887	\\						
	$E(X|Y=3) - E(X|Y=2)$     &  0.0840   &  0.0384 &  0.0705 &  0.0840 &  0.0976 	\\	
	$E(X|Y=4) - E(X|Y=3)$    &  0.0789    &  0.0421 &  0.0641 &  0.0789 &  0.0938 	\\	
\hline
\end{tabular}
\end{center}}
\end{table}
%


\section{Concluding Remarks and  Future Direction}
The primary goal of this paper is to derive the population version of Spearman's rho by using copula functions  when the  marginal distributions are discrete.  The concordance and discordance measures are applied to obtain the population version of Spearman's rho. Particularly, the probability  of ties are taken into account when discrete random variables are involved. The upper bound and lower bound of Spearman's rho with binary margins are derived which are $-0.75$ and $0.75$, respectively.  In general, since in discontinuous cases the probability of tie is positive,  the range of Spearman's rho for the discrete random variables is narrower than $[-1,1]$. 
Our theoretical and numerical results show that there is a functional relationship between Spearman's rho  abd  Kendall's tau . This  relationship is linear when the marginals are Bernoulli;  however,  it is a function of the parameters of the model when the marginals are  Binomial, Poisson, or Negative Binomial.  The maximum ratio of  Spearman's rho to  Kendall's tau  reaches to  $1.5$.
We propose and applied  a bivariate copula regression model to investigate the effect of \emph{intrauterine device} (IUD) use  on \emph{sexually transmitted diseases} (STDs) by analysing a  \emph{cervical cancer} dataset.

A natural extension of this work for future research is to consider Spearman's rho  and  Kendall's tau when one marginal is discrete and the other one is continuous.

\section*{Acknowledgement} We would like to thank the Editor in Chief, the Associate Editor, and two referees  for their helpful and constructive  comments   which led to a significant improvement of this paper.

\section*{Appendix: Proof of Theorem \ref{thr1} and Theorem \ref{thr2} }

{\bf Proof of Theorem \ref{thr1}:}
Assume $(X_{1},Y_{1})$, $(X_{2},Y_{2})$ and $(X_{3},Y_{3})$ are three independent realizations of the random vector $(X, Y)$. When $X$ and $Y$ are integer-valued random variables, we obtain $P(C)+P(D)+P(T)=1$. Subtracting the probability of discordance from both sides, we have $P(C)-P(D)=1-2P(D)-P(T)=2P(C)-1+P(T).$ Then, according to the definition of Spearman's rho in Eq \eqref{eq7} we have
\begin{align}
\rho^S (X,Y)=&3[P(C)-P(D)] \nonumber \\
=&3\{ 2P(C)-1+P(T)\} \nonumber \\
=&6\{ P[(X_{1} -X_{2} )(Y_{1} -Y_{3} )>0]\} -3+3P(X_{1} =X_{2} \, or\, Y_{1} =Y_{3} ) \nonumber\\ \label{ap.1}
=&6\{ P[X_{2} >X_{1} ,Y_{3} >Y_{1} ]+P[X_{2} <X_{1} ,Y_{3} <Y_{1} ]\} -3+3P(X_{1} =X_{2} \, \mbox{or } \, Y_{1} =Y_{3} ),
\end{align}
where,
\begin{align}\label{ap.2}
\begin{split}
P(X_{2}<X_{1},Y_{3}<Y_{1})=&\sum _{x=0}^{\infty}\sum_{y=0}^{\infty }P(X_{2}<x,Y_{3}<y)P(X_{1}=x,Y_{1} =y)\\
=&\sum _{x=0}^{\infty }\sum _{y=0}^{\infty }P(X_{2} <x)P(Y_{3} <y)  P(X_{1} =x,Y_{1} =y) \\
=&\sum _{x=0}^{\infty }\sum _{y=0}^{\infty }F(x-1)G(y-1)  h(x,y),
\end{split}
\end{align}
and similarly
\begin{align}\label{ap.3}
\begin{split}
P(X_{2}>X_{1},Y_{3}>Y_{1})=&\sum_{x=0}^{\infty }\sum_{y=0}^{\infty}P(X_{2}>x,Y_{3}>y)P(X_{1}=x,Y_{1}=y)\\
=&\sum _{x=0}^{\infty}\sum _{y=0}^{\infty }[1-F(x)][1-G(y)]h(x,y),
\end{split}
\end{align}
where $ h(x,y)$ is the joint pmf of $X$ and $Y$ and can be derived as
{\small
\begin{align*}
h(x,y)=&P(X_{1} =x,Y_{1} =y) \\
=&P(X_{1} \le x,Y_{1} \le y)-P(X_{1} \le x-1,Y_{1} \le y)
 -P(X_{1} \le x,Y_{1} \le y-1)+P(X_{1} \le x-1,Y_{1} \le y-1) \\
=&H(x,y)-H(x-1,y)-H(x,y-1)+H(x-1,y-1) \\
=&\mathcal{C}(F(x),G(y))-\mathcal{C}(F(x-1),G(y))-\mathcal{C}(F(x),G(y-1))+\mathcal{C}(F(x-1),G(y-1)).
\end{align*}}
Moreover, the last term in Eq \eqref{ap.1}  can be written as
\begin{align}\label{ap.4}
P(X_{1} =X_{2} \,  \mbox{or } \, Y_{1} =Y_{3} )=P(X_{1} =X_{2} )+P(Y_{1} =Y_{3} )-P(X_{1} =X_{2}, Y_{1} =Y_{3} ),
\end{align}
where
\begin{align}\label{ap.5}
P(X_{1} =X_{2} )&=\sum _{x=0}^{\infty }P(X_{1} =x, X_{2} =x) =\sum _{x=0}^{\infty }P(X_{1} =x)P(X_{2} =x)  =
\sum _{x=0}^{\infty }f^{2} (x),\\\label{ap.6}
P(Y_{1} =Y_{3} )&=\sum _{y=0}^{\infty }P(Y_{1} =y,  Y_{3} =y)=\sum _{y=0}^{\infty }P(Y_{1} =y)P(Y_{3}=y)=
\sum _{y=0}^{\infty }g^{2} (y),
\end{align}
and
\begin{align}\label{ap.7}
P(X_{1} =X_{2} ,Y_{1} =Y_{3} )&=\sum _{x=0}^{\infty }\sum _{y=0}^{\infty }P(X_{1} =x,Y_{1} =y)P(X_{2} =x,Y_{3} =y) \nonumber\\
&=\sum _{x=0}^{\infty }\sum _{y=0}^{\infty }P(X_{1} =x,Y_{1} =y)P(X_{2} =x)P(Y_{3} =y) \nonumber\\
&=\sum _{x=0}^{\infty }\sum _{y=0}^{\infty }h (x,y) f(x)g(y).
\end{align}
Then, by substituting the results in Eqs \eqref{ap.5}, \eqref{ap.6}, and \eqref{ap.7} into the right side of  Eq \eqref{ap.4}, we obtain
\begin{align}\label{ap.8}
P(X_{1} =X_{2} \,  \mbox{or } \, Y_{1} =Y_{3} )= \sum _{x=0}^{\infty }f^{2} (x) + \sum _{y=0}^{\infty }g^{2} (y) - \sum _{x=0}^{\infty }\sum _{y=0}^{\infty }h (x,y) f(x)g(y).
\end{align}
Finally, by substituting the expressions  \eqref{ap.7}, \eqref{ap.2}, and \eqref{ap.3} into \eqref{ap.1}, we have
\begin{align*}
\rho^S (X,Y)=&6 \sum _{x=0}^{\infty }\sum _{y=0}^{\infty }h(x,y)  \left[ (1-F(x))(1-G(y))+F(x-1)G(y-1)-\dfrac{1}{2}f(x)g(y)\right] \\
&~~~~~~~~~~~~~~~~~~~+3\sum _{x=0}^{\infty } \left(f^{2} (x)+g^{2} (x)\right) -3. \blacksquare
\end{align*}
{\bf Proof of Theorem \ref{thr2}:}
From the Bernoulli distribution, we have
\begin{align*}
 & F_{X}(-1)=G_{Y}(-1)=0,~~~~F_{X}(0)=1-p_{X},~~~~G_{Y}(0)=1-p_{Y},~~~~F_{X}(1)=G_{Y}(1)=1, \\
 & f_{X}(0)=1-p_{X},~~~~~~g_{Y}(0)=1-p_{Y},~~~~~~f_{X}(1)=p_{X},~~~~~~g_{Y}(1)=p_{Y}.
\end{align*}
 Therefore, the  Spearman's rho of two Bernoulli random variables $X$ and $Y$  can be simplified as
  \begin{align}\label{ap.9}
\rho^S (X,Y)=&6 \sum _{x=0}^{1}\sum _{y=0}^{1}h(x,y) \left[ (1-F(x))(1-G(y))+F(x-1)G(y-1)-\dfrac{1}{2}f(x)g(y)\right] \nonumber \\
 ~& ~~~~~~~~~~+3\sum _{x=0}^{1} \left(f^{2} (x)+g^{2} (x)\right) -3 \nonumber \\
=&6 h(0,0)  \big[ p_X p_Y-\dfrac{1}{2} (1-p_X)(1-p_Y)\big]  -3 h(0,1) (1-p_X)p_Y \nonumber\\
&~~~-3 h(1,0)    p_X (1-p_Y)  + 6h(1,1)  \big[(1-p_X)(1-p_Y)-\dfrac{1}{2} p_X p_Y\big] \\\nonumber
&~~~+3\left((1-p_X)^2+ (1-p_Y)^2+ p_X^2+ p_Y^2\right) -3.
\end{align}
Then, by using the fact that  $\mathcal{C}(u,0)=C(0,v)=0, \mathcal{C}(u,1)=u$, and $\mathcal{C}(1,v)=v$, all possible values of $h(x,y)$ defined in Eq \eqref{eq3.2} are obtained as follows
\begin{align*}
h(0,0)&=\mathcal{C}(F(0),G(0))-\mathcal{C}(F(-1),G(0))-\mathcal{C}(F(0),G(-1))+\mathcal{C}(F(-1),G(-1))\\
&=\mathcal{C}(1-p_{X},1-p_{Y}) - \mathcal{C}(0,1-p_{Y})-\mathcal{C}(1-p_{X},0)+\mathcal{C}(0,0)\\
&=\mathcal{C}(1-p_{X},1-p_{Y}),\\
h(0,1)
&=\mathcal{C}(1-p_{X},1) - \mathcal{C}(0,1)-\mathcal{C}(1-p_{X},1-p_{Y})+\mathcal{C}(0,1-p_{Y})\\
&=1-p_{X}-\mathcal{C}(1-p_{X},1-p_{Y}),\\
h(1,0)
&=\mathcal{C}(1,1-p_{Y}) - \mathcal{C}(1-p_{X},1-p_{Y})-\mathcal{C}(1,0)+\mathcal{C}(1-p_{X},0)\\
&=1-p_{Y}-\mathcal{C}(1-p_{X},1-p_{Y}),
\end{align*}
and
\begin{align*}
h(1,1)
&=\mathcal{C}(1,1) - \mathcal{C}(1-p_{X},1)-\mathcal{C}(1,1-p_{Y})+\mathcal{C}(1-p_{X},1-p_{Y})\\
&=p_{X}+p_{Y}+\mathcal{C}(1-p_{X},1-p_{Y})-1.
\end{align*}
Now, by substituting  the above results into the given expression of  $\rho^S (X,Y)$ in Eq \eqref{ap.9}, we obtain
\begin{align*}
\rho^S (X,Y)
=& 3 \mathcal{C}(1-p_{X} ,1-p_{Y})\left[ p_{X}p_{Y}+p_{X}+p_{Y}-1 \right] \\
&~~~-3 (1-p_{X})^2p_Y + 3 \mathcal{C}(1-p_{X} ,1-p_{Y})  \left[(1-p_{X})p_Y \right] \\
&~~~-3 p_X (1-p_{Y})^2  + 3 \mathcal{C}(1-p_{X} ,1-p_{Y})  \left[p_X(1-p_{Y})  \right] \\
&~~~+6 \left(p_{X}+p_{Y}+\mathcal{C}(1-p_{X} ,1-p_{Y})-1\right)  \left[1-p_{X}-p_{Y}+\dfrac{1}{2}p_{X}p_{Y}\right] \\
&~~~+3\left(2-2p_{X}-2p_{Y}+2p^{2}_{X}+2p^{2}_{Y}\right) -3\\
=&-3+3\mathcal{C}(1-p_{X} ,1-p_{Y})+3p_{X}+3p_{Y}-3p_{X}p_{Y},
\end{align*}
 and the proof is completed.  $~~~ \blacksquare$



\end{document}